\documentclass[a4paper,10pt]{article}

\usepackage[utf8]{inputenc}
\usepackage {graphicx}
\usepackage {amsfonts}
\usepackage{amsmath}
\usepackage{url}
\usepackage{hyperref}
\usepackage{amsthm}
\usepackage{mathtools}
\usepackage{bbm}
\usepackage{dsfont}
\usepackage{textcomp}
\usepackage{natbib}
\usepackage[colorinlistoftodos]{todonotes}
\usepackage{algorithm,algorithmic}
\usepackage{booktabs}

\makeatletter

\graphicspath{{./pics/}}

\newtheorem{theorem}{Theorem}[section]
\newtheorem{definition}[theorem]{Definition}

\newtheorem{scheme}[]{Scheme}
\newtheorem*{teorema}{Theorem}

\newcommand{\numberset}{\mathbb}
\newcommand{\N}{\numberset{N}}
\newcommand{\R}{\numberset{R}}
\newcommand{\E}{\numberset{E}}

\newcommand{\PP}{\numberset{P}}
\newcommand{\Q}{\numberset{Q}}
\newcommand{\SI}{\mathcal{S}}
\newcommand{\LL}{\mathcal{L}}
\newcommand{\xx}{\mathrm{x}}

\newcommand{\footremember}[2]{%
    \footnote{#2}
    \newcounter{#1}
    \setcounter{#1}{\value{footnote}}%
}

\title{Option pricing in exponential Lévy models with transaction costs.}

\author{ Nicola Cantarutti$^\dagger$\footremember{alley}{Corresponding author. Email: \href{mailto:nicolacantarutti@gmail.com}{\tt nicolacantarutti@gmail.com}} ,
 Manuel Guerra$^{\dagger}$, João Guerra$^{\dagger}$,\\ Maria do Rosário Grossinho$^{\dagger}$ \\
 \vspace{0.5em} \\
\small $^\dagger$CEMAPRE - Center for Applied Mathematics and Economics\\ \small ISEG - University of Lisbon }

\begin{document}

\maketitle

\begin{abstract}
 We present an approach for pricing European call options in presence of proportional transaction costs,
 when the stock price follows a general exponential L\'{e}vy process. 
 The model is a generalization of the celebrated work of Davis, Panas and Zariphopoulou (1993), 
 where the value of the option is defined as
 the utility indifference price. This approach requires the solution of
 two stochastic singular control problems in finite horizon, satisfying the same
 Hamilton-Jacobi-Bellman equation, with different terminal conditions.
 We introduce a general formulation for these portfolio selection problems, and then we
 focus on the special case in which the probability of default is ignored.
 We solve numerically the optimization problems using the Markov chain approximation method and
 show results for diffusion, Merton and Variance Gamma processes.
 Option prices are computed for both the writer and the buyer.  

 \vspace{1em}
\end{abstract}

\section{Introduction}

The problem of pricing a European call option was first solved mathematically in the paper \cite{BS73}. 
Even if it is quite evident that this model is too simplistic to represent the real features of the market, it is 
still nowadays one of the most used models to price and hedge options.
The reason for its success is that it gives a closed form solution for the option price, and that the hedging strategy is easily 
implementable.
One of the main assumptions of the Black-Scholes model is that
log-returns are normally distributed. 
However, statistical analysis of financial data
reveal that the normality assumption is not a very good approximation of
reality (see for instance \cite{Cont01}). Empirical log-return distributions have
more mass around the origin and along the tails (\emph{heavy tails}).
This means that the normal distribution underestimates the probability of large
log-returns, and considers them just as rare events. In the real market instead,
log-returns manifest frequently high peaks, that come more and more evident
when looking at short time scales. The log-returns peaks correspond to sudden
large changes in the price. 
There is a huge literature of option pricing models that consider underlying processes with discontinuous paths.
Most of these models consider the log-price dynamics following a \emph{Lévy process} i.e. a 
stochastic process with independent and stationary increments,
satisfying the additional property of stochastic continuity.
Good references on the theory of Lévy processes are the books of \cite{Sato} and \cite{Applebaum}. 
Financial applications are discussed in the book of \cite{Cont}.  

Another issue of the Black-Scholes framework is that it does not consider the presence of budget constraints and
market frictions such as bid/ask spread and transaction fees.
In particular, in a market with transaction costs the replicating portfolio cannot be perfectly implemented.
Since delta-hedging involves continuous time trading, 
the presence of transaction costs makes this strategy infinitely costly.
Many authors attempted to include proportional transaction costs in option pricing models.
In \cite{Le85}, in order to avoid continuous trading, the author specifies 
a finite number of trading dates. He obtains a Black-Scholes-like
nonlinear partial differential equation (PDE) with an adjusted volatility term that takes into account the transaction costs. 
However this model has several drawbacks, i.e. trading at fixed dates is not optimal and the option price diverges as the number of trading times grows. 
Recent developments in this direction are for instance \cite{Mocio07}, \cite{FlMaSe14} and \cite{Sengu14} 
who consider different features of the market such as jumps, stochastic volatility and stochastic interest rate respectively.  

A different approach has been introduced by \cite{HoNe89}. The authors use an alternative definition of the option price
called \emph{indifference price}, based on the \emph{expected utility maximization} concept.  
An overview of these topics applied to several incomplete market models can be found in \cite{Carmona}.
Since a perfect replicating portfolio does not exist, there is no risk-free hedging strategy.
The model has to take into account the risk profile of the writer/buyer to describe his trading preferences.
\cite{HoNe89} define the option price as the value that makes an investor indifferent between holding a portfolio with an option
and holding a portfolio with this additional value, in terms of expected utility of the final wealth.
The optimal hedging strategy is to keep the portfolio within
a band called \emph{No Transaction} region. Using numerical experiments, they verify that this strategy outperforms the one 
proposed in \cite{Le85}.
This approach has been further developed in \cite{DaPaZa93}, where the problem is formulated rigorously as a singular 
stochastic optimal control problem. The authors prove that the value function of the optimization problem
can be interpreted as the solution of the associated Hamilton-Jacobi-Bellman (HJB) equation in the viscosity sense. 
They prove also that the numerical scheme, based on the \emph{Markov chain approximation}, converges to the viscosity solution.
Numerical methods for this model are presented in \cite{DaPa94}, \cite{ClHo97} and \cite{Mon03}, \cite{Mon04}.
In \cite{WhWi97} and \cite{BaSo98} the problem is simplified by using asymptotic analysis for small 
transaction costs, in order to reduce the original variational inequality to a simpler non-linear PDE.
Further developments are presented in the thesis work of \cite{Damgaard}, where the author 
studies the robustness of the model with respect to the choice of the utility function.  

In the present work, we want to develop a model for pricing options using the concept of indifference price proposed in \cite{HoNe89} under
the presence of proportional transaction costs. We consider the stock dynamics evolving as an exponential Lévy process.
Portfolio models with Lévy processes and transaction costs have already appeared in the literature. For instance, in \cite{Benth02} and \cite{OkSu01}, 
the authors, in order to exclude the possibility of bankruptcy, introduce strong restrictions in the set of admissible trading strategies, i.e. short-selling 
and borrowing money are not allowed.
\cite{Kab16} formulate a more general framework, where negative positions in stocks and cash are allowed, and bankruptcy is considered. 
They prove that the infinite horizon singular/regular control problem considered in their article, satisfies the dynamic programming principle and show that the value function 
is the unique continuous viscosity solution of the HJB equation.

The model we present in Section \ref{model} is an adaptation of the model of \cite{Kab16} to our finite horizon problem.
Since the numerical methods for this problem are quite cumbersome,
we focus on the special case of an investor with a very large \emph{credit availability} i.e. we ignore the possibility of default. 
By assuming a risk profile described by an exponential utility function, we reduce by one 
the number of state variables and obtain a simpler optimization problem.
In Section \ref{MC_section} we present the discretized version of the problem.
We propose a monotone, stable and consistent numerical scheme and prove that its solution converges to the viscosity solution of the HJB equation.
The discretization technique is described in the Appendix \ref{B}, while the Appendix \ref{C} contains the proofs of the theorems.
The numerical results are presented in Section \ref{numerical}. 
Section \ref{conclusions} contains a summary of the outcomes, with suggestions for future improvements.

\section{The model}\label{model}

\subsection{Exponential Lévy models}
Let $\{X_t\}_{t \in [t_0,T]}$ be a Lévy process defined on the filtered probability space $\bigl( \Omega,\mathcal{F},\{\mathcal{F}_{t}\}_{t \in [t_0,T]},\PP \bigr)$, 
where $\{\mathcal{F}_{t}\}_{t \in [t_0,T]}$ is the natural filtration.
We assume that $\{X_t\}_{t \in [t_0,T]}$ has the characteristic Lévy triplet $(b, \sigma, \nu)$, where $b \in \R$, $\sigma \geq 0$ and $\nu$ 
is a positive measure on $\R$, called \emph{Lévy measure} which satisfies:
\begin{equation} \label{Levy_m}
 \nu \bigl(\{ 0 \} \bigr) = 0, \hspace{2em}
 \int_{\R} \bigl( 1 \wedge z^2 \bigr) \nu(dz) < \infty.
\end{equation}
We use the (cádlág) process $\{X_t\}_{t \in [t_0,T]}$ to model the log-price dynamics.
The stock price $\{S_t\}_{t \in [t_0,T]}$ is therefore an \emph{exponential Lévy process} such that: 
\begin{equation}\label{ELM}
 S_t = S_0 e^{X_t}.
\end{equation}
Motivated by practical reasons, we only consider processes with finite mean and variance.
The condition for a finite second moment 
$\E[S_t^2] < \infty$,
is directly related to the integrability conditions of the Lévy measure: 
\begin{equation}\label{fin_moment}
\int_{|z| \geq 1} e^{2z} \nu(dz) <\infty. 
\end{equation}
Considering condition (\ref{fin_moment}), the dynamics of $\{X_t\}_{t \in [t_0,T]}$ has the following \emph{Lévy-Itô decomposition}:
\begin{equation}\label{log_sde} 
 dX_t = \biggl( b + \int_{|z|\geq 1}z \nu(dz) \biggr) \,dt + \sigma dW_t + \int_{\R} z \tilde N(dt,dz),
\end{equation}
where $W_t$ is a standard Brownian motion.
The term 
\begin{equation}
 \tilde N(dt,dz) := N(dt,dz) - dt\, \nu(dz), 
\end{equation}
is the compensated Poisson martingale measure, 
where $N(dt,dz)$ is the Poisson random measure with intensity $dt \, \nu(dz)$.
By applying the Itô lemma to (\ref{ELM}) we obtain the SDE describing the evolution of the price: 
\begin{equation}\label{exp_sde}
 \frac{d S_t}{S_{t^-}} = \; \mu \, dt +  \sigma dW_t \; + \int_{\R} (e^{z} - 1) \tilde N(dt,dz), 
\end{equation}
where $S_{t^-} := \lim_{u\uparrow t}S_{u}$.
We defined the drift term as: 
\begin{equation}\label{mu}
\mu :=  b + \frac{1}{2}\sigma^2 + \int_{\R} \bigl( e^{z} - 1 -z\mathbbm{1}_{\{ |z|<1 \}} \bigr) \nu(dz). 
\end{equation}

\subsection{Portfolio dynamics with transaction costs}

In this section we follow the framework of \cite{Kab16} and introduce a market model with proportional transaction costs that generalizes \cite{DaPaZa93}. 
Let us consider a portfolio composed by one risk-free asset $B$ (bank account)
paying a fixed interest rate $r > 0$ and a stock $S$. 
We denote by $Y$ the number of shares of the stock $S$ that the investor holds.
The state of the portfolio at time $t\in [t_0,T]$ is $(B^{\pi}_t,Y^{\pi}_t,S_t)$ and evolves following the SDE:
\begin{equation}\label{porfolio_dynamics}
 \begin{cases}
 dB^{\pi}_t &=  rB^{\pi}_t dt - (1+\theta_b)S_{t^-} dL_t + (1-\theta_s) S_{t^-} dM_t \\
 dY^{\pi}_t &=  dL_t - dM_t \\
 dS_t &=  S_{t^-} \left( \mu dt + \sigma dW_t + \int_{\R} (e^z-1) \tilde N(dt,dz) \right).
\end{cases}
\end{equation} 
The parameters $\theta_b$, $\theta_s \geq 0$ are the proportional transaction costs when buying and selling respectively. 
The control process $\{\pi_t\}_{t \in [t_0,T]} := \{(L_t,M_t)\}_{t \in [t_0,T]}$ is the trading strategy that represents the 
cumulative number of shares respectively bought and sold up to time $t$.
The strategy $\{\pi_t\}_{t \in [t_0,T]}$ is a cádlág, predictable, nondecreasing process with bounded variation, such that
$ \pi(t_0^-) = \bigl ( L(t_0^-) , M(t_0^-) \bigr ) = (0,0)$, i.e. we allow for an initial transaction.
Under these assumptions the portfolio process 
$\bigl \{(B^{\pi}_t,Y^{\pi}_t,S_t)\bigr \}_{t \in [t_0,T]}$ is cádlág.

If at time $t$ there is an unpredictable jump in the stock price $\Delta S_t = S_t - S_{t^-}$, a possible transaction should happen immediately after the jump\footnote{
The control process $\{\pi_t\}_{t \in [t_0,T]}$ is assumed to be predictable, i.e. measurable with respect to the left-continuous filtration generated by $\{S_{t^-}\}_{t \in [t_0,T]}$.
Therefore, a jump in the price and a jump in the control cannot occur simultaneously, almost surely. 
A deeper digression on this topic can be found in Section 2 of \cite{Kab16}.}.   
If the investor at time $t$ observes a jump in the price and decides to rebalance his portfolio, he will trade at some time $u>t$ at the price $S_{u^-}$. 
Under this framework, as explained in \cite{Kab16}, the optimal strategy cannot exist.    

\begin{definition}
The \textbf{cash value} function $c(y,s) : \R \times \R^+ \to \R$, is defined as the value in cash when the shares in the portfolio are liquidated i.e.  
long positions are sold and short positions are covered.
\begin{equation}\label{cost_function}
c(y,s) := \begin{cases} 
(1+\theta_b)ys, & \mbox{if } y\leq 0 \\ 
(1-\theta_s)ys, & \mbox{if } y>0 . 
\end{cases} 
\end{equation}
\end{definition}
\begin{definition}
For $t\in [t_0,T]$, the \textbf{total wealth} process is defined as:
\begin{equation}\label{wealth_process}
 \mathcal{W}^{\pi}_t := B^{\pi}_t + c(Y^{\pi}_t,S_t).
\end{equation} 
\end{definition}
We say that a portfolio is solvent at time $t$ if the portfolio's wealth $ \mathcal{W}^{\pi}_t $ is bigger than a fixed constant $-C$, with 
$C \geq 0$. This constant may depend on the initial wealth and on the parameters in (\ref{porfolio_dynamics}). 
It can be interpreted as the \textbf{credit availability} of the investor.
\begin{definition}
 The \textbf{solvency region} is defined as:
\begin{equation}\label{solvency_region}
 \mathcal{S} := \biggl\{ (b,y,s) \in \R \times \R \times \R^+ : b + c(y,s) > -C  \biggr\}.
\end{equation}
\end{definition}
Since the underlying stock follows a process with jumps, it is not guaranteed that the portfolio stays
solvent for all $t \in [t_0,T]$. When holding short positions, it is possible that a sudden increase in the stock price 
causes the total wealth to jump out of the solvency region. 
The same can happen with a downward jump when the investor is long in stocks and negative in cash. 
An immediate decrease in the stock price would make him unable to pay the debts.
If the investor goes bankrupt, there are no trading strategies to save him.
\begin{definition}
The first \textbf{exit time} from the solvency region is defined as:
\begin{equation}\label{exit_time}
 \tau := \inf \bigl\{ t \in [t_0,T] : \mathcal{W}^{\pi}_t \not\in \SI \bigr\}.
\end{equation} 
\end{definition}
\begin{definition}\label{set_trad_strat}
The set of \textbf{admissible trading strategies} $\Pi(t_0,b,y,s)$   
is the set of all cádlág, nondecreasing, predictable, bounded variation processes $\{\pi_t\}_{t \in [t_0,T]}$
such that $(B^\pi_t,Y^\pi_t,S_t)$ is a solution of (\ref{porfolio_dynamics}) with initial values $(B^\pi_{t_0} = b, Y^\pi_{t_0} = y, S_{t_0} = s)$ and such that
$\mathcal{W}^{\pi}_t \in \SI$ almost surely for all $t\in [t_0,\tau)$, and $\pi_t = \pi_{\tau}$ for all $t \geq \tau$.  
\end{definition}

\subsection{Utility maximization}

Let us assume an investor builds a portfolio with cash, shares of a stock and in addition he sells or purchases a 
European call option written on the same stock, with strike price $K$ and expiration date $T$.\\
From now on, we introduce the superscripts $w$, $b$ and $0$ to indicate the writer, buyer and zero-option portfolios respectively.
In the zero-option portfolio, the wealth process $\{\mathcal{W}^{0; \pi}_t\}_{t \in [t_0,T]}$ and the exit time $\tau^0$ correspond to (\ref{wealth_process}) and (\ref{exit_time}).
\begin{definition}
For $t \in [t_0,T]$, we define the wealth processes for the writer:
  \begin{align}\label{wealth_writer}
   \mathcal{W}^{w; \pi}_t := & \; B^{\pi}_t + c(Y^{\pi}_t,S_t) \mathbbm{1}_{\{t < T\}} \\ \nonumber
   & + c(Y^{\pi}_t,S_t) \mathbbm{1}_{\{t = T,\, S_t(1+ \theta_b ) \leq K\}}
   + \biggl( c\bigl( Y^{\pi}_t-1,S_t \bigr) + K \biggr) \mathbbm{1}_{\{t=T,\, S_t(1+ \theta_b ) > K \}}
  \end{align}
 and the buyer:
  \begin{align}\label{wealth_buyer}
   \mathcal{W}^{b; \pi}_t := & \; B^{\pi}_t + c(Y^{\pi}_t,S_t) \mathbbm{1}_{\{t < T\}} \\ \nonumber
   & + c(Y^{\pi}_t,S_t) \mathbbm{1}_{\{t = T,\, S_t(1+ \theta_b ) \leq K\}}
   + \biggl( c\bigl( Y^{\pi}_t+1,S_t \bigr) - K \biggr) \mathbbm{1}_{\{t=T,\, S_t(1+ \theta_b ) > K \}}.
  \end{align}
\end{definition}
In the case the option is exercised, $S_T(1+ \theta_b ) > K$, the buyer pays to the writer the strike value $K$ in cash, 
and the writer delivers one share to the buyer.
In a market with transaction costs, the real value (in cash) of a share incorporates the transaction cost for the purchase or the sale. 
Therefore the buyer does not exercise when $S_T > K$, but when $S_T(1+ \theta_b ) > K$.

The objective of the investor is to maximize the expected utility of the wealth at $\tau^j \wedge T$ over all the admissible
strategies. This expectation is conditioned on the current amount of cash and number of shares in the portfolio and on the current stock price.
The exit times $\tau^w$ and $\tau^b$ for the writer and buyer portfolios, can be obtained by inserting (\ref{wealth_writer}) and (\ref{wealth_buyer}) in (\ref{exit_time}). 
The sets of trading strategies for the three portfolios can be obtained in the same way from the Definition [\ref{set_trad_strat}]. 

\begin{definition}
 The \textbf{value function} of the maximization problem for $j=w,b,0$ is defined as:
\begin{align}\label{max_probl1}
V^j(t,b,y,s) := \sup_{\pi \in \Pi^j(t,b,y,s)} \; & \E_{t,b,y,s}\biggl[ 
            \mathcal{U}(\mathcal{W}^{j; \pi}_T) \; \mathbbm{1}_{\{\tau^j > T\}} \\ \nonumber
             &+ e^{ \beta (T-\tau^j)} \mathcal{U}( \mathcal{W}^{j; \pi}_{\tau^j} ) \; 
             \mathbbm{1}_{\{\tau^j \leq T\}}\biggr],             
\end{align}
where $\mathcal{U}: \R \to \R$ is a concave increasing utility function such that $\mathcal{U}(0)=0$, and $\beta \geq 0$.
\end{definition}

The option price for the writer (buyer), is defined as the amount of cash to add (subtract) to the bank account 
of the portfolio containing the option, such that the maximal expected utility of final wealth of the writer (buyer) is the same he could get with 
the zero-option portfolio.
\begin{definition}
The writer price is the value $p^w>0$ such that 
 \begin{equation}\label{writer}
  V^0(t,b,y,s) = V^w(t,b+p^w,y,s),
 \end{equation}
 and the buyer price is the value $p^b>0$ such that
 \begin{equation}\label{buyer}
  V^0(t,b,y,s) = V^b(t,b-p^b,y,s).
 \end{equation}
\end{definition}

\subsection{Hamilton-Jacobi-Bellman Equation}

We present the HJB equation associated to the singular stochastic optimal control problem described before.
This problem is called singular
because the controls $\{(L_t,M_t)\}_{t \in [t_0,T]}$ are allowed to be singular with
respect to the Lebesgue measure $dt$. 
The derivation of the HJB equation for singular control problems can be found for instance in \cite{FlemingSoner}.  
We assume without proving that the value function (\ref{max_probl1}) satisfies the dynamic programming principle.
The HJB equation associated to (\ref{max_probl1}) is the following variational inequality:
\begin{align}\label{HJB1}
& \max \; \biggl\{ \; \frac{\partial V^j}{\partial t} + rb\frac{\partial V^j}{\partial b} 
+ \mu s \frac{\partial V^j}{\partial s} + \frac{1}{2}\sigma^2 s^2 \frac{\partial^2 V^j}{\partial s^2} \\ \nonumber
&+ \int_\mathbb{R}
\biggl[ V^j(t,b,y,se^z) - V^j(t,b,y,s) - s(e^z-1)\frac{\partial V^j}{\partial s} \biggr] \nu(dz) \;,\\ \nonumber
& \;  \frac{\partial V^j}{\partial y}-(1+\theta_b) s \frac{\partial V^j}{\partial b} \; 
, \; -\biggl(\frac{\partial V^j}{\partial y}-(1-\theta_s)s \frac{\partial V^j}{\partial b} \biggr) \biggr\} = 0, 
 \end{align}
for $(t,b,y,s) \in [t_0,T] \times \SI$ and $j=0,w,b$.
The terminal boundary conditions are given by Eq. (\ref{max_probl1}) at time $T$.
Since this HJB equation is a PIDE,
the non-local integral operator implies the
definition of lateral conditions not only on the boundaries
of the solvency region, but also beyond:
\begin{equation}\label{lat_bound}
V^j(t,b,y,s) = \mathcal{U}\bigl( b + c(y,s)\bigr) \hspace{1em} \mbox{ for } \hspace{1em} 
t \in [t_0,T) , \hspace{0.5em} (b,y,s) \not \in \mathcal{S}, \hspace{1em} j=0,w,b. 
\end{equation}
We refer to the arguments in \cite{Kab16} to prove that the value function (\ref{max_probl1}) is the unique continuous viscosity solution of (\ref{HJB1}).
 

\subsection{Variable reduction}\label{Section2.5}

In the diffusion model of \cite{DaPaZa93}, the portfolio is solvent for every
$t \in [t_0,T]$ (almost surely) and it is always possible to calculate the utility 
of the wealth at the terminal time $\mathcal{U}(\mathcal{W}^{\pi}_T)$.
In the framework of \cite{Kab16} the stock process can jump, and in presence of short positions the portfolio can go bankrupt at any time before the maturity $T$. 

With the intention of simplifying the maximization problem (\ref{max_probl1}) and reducing the number of variables, 
we restrict our attention to the case of no bankruptcy.
A possible idea is to consider a positive initial wealth, and define the restricted set of admissible strategies as the set of $\{\pi_t\}_{t \in [t_0,T]}$ such that 
$B^{\pi}_t \geq 0$ and $Y^{\pi}_t \geq 0$ for all $t\in [t_0,T]$ (see \cite{Benth02}).
However, in order to implement a hedging strategy, we are interested in portfolios containing short positions as well.
So, we can assume that the investor has a very large credit availability $C$ in the sense that
\begin{equation}\label{P_tau}
 \PP(\tau > T) \underset{C\to \infty}{\approx} 1.
\end{equation}
In practical terms, we ignore the possibility of default. The solvency region becomes $\mathcal{S} = \R^2 \times \R^{+}$ and no lateral boundary conditions are imposed.

As in \cite{DaPaZa93}, for $\gamma>0$, we consider the exponential utility function
\begin{equation}\label{exp_util}
 \mathcal{U}(x) := 1- e^{-\gamma x}.
\end{equation}
Thanks to (\ref{P_tau}) and (\ref{exp_util}) we can remove $\{B^{\pi}_t\}_{t \in [t_0,T]}$ from the state dynamics.
By solving (\ref{porfolio_dynamics}) we get
\begin{equation}\label{BT}
B^{\pi}_T =  \frac{B^{\pi}_{t}}{\delta(t,T)} - \int_{t}^T
(1+\theta_b)\frac{S_u}{\delta(u,T)} dL_u + \int_{t}^T
 (1-\theta_s) \frac{S_u}{\delta(u,T)} dM_u 
\end{equation}
where $\delta(u,T) = e^{-r(T-u)}$.
Using together (\ref{P_tau}), (\ref{exp_util}) and (\ref{BT}), and the wealth processes (\ref{wealth_process}),(\ref{wealth_writer}),(\ref{wealth_buyer}), 
we obtain for $B^\pi_{t} = b$, $Y^\pi_{t} = y$, $S_{t} = s$ and $j=0,w,b$:
\begin{align}\label{var_reduct}
   V^j(t,b,y,s) = \sup_{\pi} \; \E_{t,b,y,s}\biggl[  1- e^{-\gamma \mathcal{W}^j(T) } \biggr]  
	     = 1- e^{-\gamma \frac{b}{\delta(t,T)}} Q^j(t,y,s),
\end{align} 
where
\begin{align}\label{minimization}
Q^j(t,y,s) = \inf_{\pi} \; \mathbb{E}_{t,y,s}\biggl[ \; &
	     e^{-\gamma \bigl[ -\int_{t}^T (1+\theta_b) \frac{S_u}{\delta(u,T)} dL_u +
	     \int_{t}^T (1-\theta_s) \frac{S_u}{\delta(u,T)} dM_u \bigr] } \, \\ \nonumber 
	     & \times H^j(Y^{\pi}_T,S_T) \bigg]  
\end{align}
is our new minimization problem.
The exponential term inside the expectation can be considered as a discount factor, and the second term 
$H^j(y,s) = Q^j(T,y,s)$ is the terminal payoff:
\begin{itemize}
 \item No option:
 \begin{equation}\label{terminal_c}
  H^0(y,s) = e^{-\gamma \, c(y,s)}.
 \end{equation}
 \item Writer:
  \begin{equation}\label{terminal_w}
  H^w(y,s) = e^{-\gamma \bigl[ c(y,s)\mathbbm{1}_{\{s(1+\theta_b) \leq K\}} + 
 \bigl( c( y-1,s) + K \bigr) \mathbbm{1}_{\{s(1+\theta_b)>K\}} \bigr] }.
 \end{equation}
 \item Buyer:
  \begin{equation}\label{terminal_b}
  H^b(y,s) = e^{-\gamma \bigl[ c(y,s)\mathbbm{1}_{\{s(1+\theta_b) \leq K\}} + 
 \bigl( c( y+1,s) - K \bigr) \mathbbm{1}_{\{s(1+\theta_b)>K\}} \bigr] }.
 \end{equation}
\end{itemize}
Using conditions (\ref{writer}), (\ref{buyer}) together with (\ref{var_reduct}), we obtain the explicit formulas for the option prices:
\begin{equation}\label{opt_w}
 p^w(t_0,y,s) = \frac{\delta(t_0,T)}{\gamma} \log \biggl( \frac{Q^w(t_0,y,s)}{Q^0(t_0,y,s)} \biggr),
\end{equation}
\begin{equation}\label{opt_b}
 p^b(t_0,y,s) = \frac{\delta(t_0,T)}{\gamma} \log \biggl( \frac{Q^0(t_0,y,s)}{Q^b(t_0,y,s)} \biggr).
\end{equation}

Since $Q^j(t,y,s)$ is independent on $b$, let us write
$ Q^j(t,y,s) := 1 - V^j(t,0,y,s)$.
It is convenient to pass to the log-variable $x = \log(s)$, such that
\begin{equation}\label{log_var}
s \frac{\partial}{\partial s} = \frac{\partial}{\partial x}, \hspace{2em} 
s^2 \frac{\partial^2}{\partial s^2} = \frac{\partial^2}{\partial x^2} - \frac{\partial}{\partial x} . 
\end{equation}
For $j=0,w,b$, the HJB Eq. (\ref{HJB1}) becomes:
\begin{align}\label{HJB2}
& \min \; \biggl\{ \; \frac{\partial Q^j}{\partial t} + (\mu-\frac{1}{2}\sigma^2) \frac{\partial Q^j}{\partial x}
+ \frac{1}{2}\sigma^2 \frac{\partial^2 Q^j}{\partial x^2} \\ \nonumber
&+ \int_\mathbb{R}
\biggl[ Q^j(t,y,x+z) - Q^j(t,y,x) - (e^z-1)\frac{\partial Q^j}{\partial x} \biggr] \nu(dz) \;,  \\ \nonumber
& \; \frac{\partial Q^j}{\partial y} +(1+\theta_b) e^x \frac{\gamma}{\delta(t,T)}Q^j \; , 
\; -\biggl( \frac{\partial Q^j}{\partial y}+(1-\theta_s)e^x \frac{\gamma}{\delta(t,T)} Q^j 
\biggr) \biggr\} = 0. 
 \end{align}
This equation is well defined for $Q^j \in C^{1,1,2}\bigl( [t_0,T] \times \R^2\bigr) \bigcap C_2\bigl( [t_0,T] \times \R^2 \bigr) $, 
where $C_2\bigl( [t_0,T] \times \R^2 \bigr)$ is the space
of continuous functions with quadratic polynomial growth at infinity in $x \in \R$, for each $t,y \in [t_0,T] \times \R$.
Since in (\ref{minimization}) we do not make any assumptions on the smoothness of $Q^j$, we will assume that it satisfies the Equation (\ref{HJB2}) in the viscosity sense. 
In the following, we will also assume that $Q^j$ is continuous (this hypothesis follows the argument in Section 4 of \cite{Kab16}). 


\section{Markov chain approximation} \label{MC_section}

To solve the minimization problem (\ref{minimization}) we use the Markov chain approximation method described in \cite{Kushner}.
The numerical technique specific for singular controls has been developed in \cite{MaKu91}.
The portfolio dynamics (\ref{porfolio_dynamics}) is approximated by a discrete state controlled Markov chain in discrete time. 
The method consists in creating a backward recursive 
dynamic programming algorithm, in order to compute the value function at time $t$, given its value at time $t+\Delta t$.
\cite{Kushner} prove that the value function obtained through the discrete dynamic programming algorithm converges to 
the value function of the original problem using a ``convergence in probability'' argument.
In \cite{BaSo91}, the authors consider instead the convergence of the discrete value function to the viscosity solution of the HJB equation.
\cite{DaPaZa93} prove the existence and uniqueness of the viscosity solution
of the HJB Eq. (\ref{HJB1}) for the diffusion case, and using the method developed by \cite{BaSo91} prove 
that the discrete value function, obtained through the Markov chain approximation, converges to it.
In this section we propose a discretization scheme and prove that it is monotone, consistent, stable, and its solution converges to the continuous viscosity
solution of (\ref{HJB2}).

In this work we model the stock dynamics with a general exponential Lévy process. In practical computations, we need to specify
which Lévy process we are using, and this is equivalent to select a Lévy triplet.
Since every L\'evy process satisfies the Markov property, we are allowed to use the Markov chain approximation approach.  
A possible way to construct the Markov chain is to discretize the infinitesimal generator by using an explicit finite difference method
(see for instance \cite{Kushner} or \cite{FlemingSoner}).
This is straightforward for 
Lévy processes of jump-diffusion type with finite jump activity. 
But for Lévy processes with infinite jump activity, it is not straightforward to obtain the transition probabilities from the discretization of the generator.
A common procedure is to approximate the small jumps with a Brownian motion, as explained in \cite{CoVo05b}, in order to remove
the singularity of the Lévy measure near the origin.

\subsection{The discrete model}\label{discrete_model}

Thanks to the variable reduction introduced in the previous section, the optimization problem (\ref{minimization}) only depends on two state variables. 
The portfolio dynamics (\ref{porfolio_dynamics}) has the simpler form (using $X_t = \log S_t$):
\begin{equation}\label{portfolio_dynamics2}
 \begin{cases}
 dY^{\pi}_t &=  dL_t - dM_t \\
 dX_t &= \biggl( \mu - \frac{1}{2} \sigma^2 - \int_{\R} (e^z-1-z) \nu(dz) \biggr) dt + \sigma dW_t + \int_{\R} z \tilde N (dt,dz).
\end{cases}
\end{equation} 
where the SDE for the log-variable can be obtained by putting together (\ref{log_sde}) and (\ref{mu}).
If the process has finite activity $\lambda := \int_{\R} \nu(dz)$, thanks to the moment condition (\ref{fin_moment}),
we can define $m := \int_{\R} \bigl( e^z -1 \bigr) \nu(dz)$ and $\alpha := \frac{1}{\lambda} \int_{\R} z \nu(dz)$ such that the SDE of $\{X_t\}_{t \in [t_0,T]}$ can be written as 
\begin{equation}\label{log_sde_Merton} 
 dX_t = \biggl( \mu - \frac{1}{2}\sigma^2 -m + \lambda \alpha \biggr) \,dt + \sigma dW_t + \int_{\R} z \tilde N(dt,dz). \\
\end{equation}
If the process has infinite activity $\int_{\R} \nu(dz) = \infty$, 
we can approximate the ``small jumps'' martingale component with a Brownian motion with same variance.
After fixing a truncation parameter $\epsilon >0$, we can split the integrals in (\ref{portfolio_dynamics2}) in two domains $\{|z|<\epsilon\}$ and $\{|z|\geq \epsilon\}$.
The integrand on the domain $\{ |z|<\epsilon \}$, is approximated by Taylor expansion 
 $e^z-1-z = \frac{z^2}{2} + \mathcal{O}(z^3)$ such that:
\begin{align}\label{log_sde_inf_act}\nonumber
  dX_t =& \biggl( \mu - \frac{1}{2}\sigma^2 -\int_{|z|<\epsilon} (e^z-1-z) \nu(dz) -\int_{|z|\geq \epsilon} (e^z-1-z) \nu(dz)  \biggr) dt\\ \nonumber
        &+ \sigma dW_t + \underbrace{\int_{|z|< \epsilon} z \tilde N(dt,dz)}_{\approx \sigma_{\epsilon} dW_t} + \int_{|z| \geq \epsilon} z \tilde N(dt,dz) \\ 
       =& \biggl( \mu - \frac{1}{2} (\sigma^2 + \sigma_{\epsilon}^2) - \omega_{\epsilon} + \lambda_{\epsilon} \theta_{\epsilon}  \biggr) dt + \bigl( \sigma+\sigma_{\epsilon}\bigr) dW_t 
       + \int_{|z|\geq \epsilon} z \tilde N(dt,dz) ,
\end{align}
where we defined the new parameters: 
\begin{align}\label{sig_eps}
 & \sigma_{\epsilon}^2 :=  \int_{|z| < \epsilon} z^2 \nu(dz), \quad \quad \omega_{\epsilon} := \int_{|z| \geq \epsilon} (e^z-1) \nu(dz), \\ \nonumber
 & \lambda_{\epsilon} :=  \int_{|z| \geq \epsilon} \nu(dz), \quad \quad \theta_{\epsilon} := \frac{1}{\lambda_{\epsilon}} \int_{|z| \geq \epsilon} z \nu(dz) .
\end{align}
The process $\int_{|z|\geq \epsilon} z \tilde N(dt,dz)$ is a compensated Poisson process with finite activity $\lambda_{\epsilon}$ 
and $\sigma_J^2 := \int_{|z| \geq \epsilon} z^2 \nu(dz) $.

Now we can discretize the time and space to create a Markov chain approximation of the portfolio process (\ref{portfolio_dynamics2}).
For $n = 0,1, ... N \in \N$, we define the discrete time step $ \Delta t := \frac{T - t_0}{N} $ such that
$t_n = t_0 + n \Delta t$.
We assume that the controls $\bigl(L_u,M_u\bigr)$ are constant for $u \in [t_n,t_{n+1})$, and allow for a possible variation at $t_n$ for each $n$.
From now on, we indicate $X_n := X(t_n)$ and the process $Y_n := Y(t^-_n)$ immediately before the possible transaction. 

Let us define the set $\Sigma_x := \{-K_1 h_x , ... , -h_x,0,h_x, ... , +K_2 h_x \}$,  
where $h_x>0$ is the discrete log-return step. The values $K_1,K_2 \in \N$ can be 
different to capture the possible asymmetry in the jump sizes. Its dimension is $\bar L = \#(\Sigma_x) = K_1 + K_2 +1$.
Let us define also the set $\Sigma_y := \{-K_3 h_y , ... , -h_y,0,h_y, ... , + K_4 h_y \} $, 
where $h_y>0$ is the discrete shares step and $K_3,K_4 \in \N$. Its dimension is $\bar M = \#(\Sigma_y) = K_3+K_4+1$.
The discretized version of the SDE (\ref{portfolio_dynamics2}) is: 
\begin{equation}\label{log_sde_discr}
 \begin{cases}
 \Delta Y_n &= \; \Delta L_n - \Delta M_n \\
 \Delta X_n &= \; \hat \mu  \Delta t + \hat \sigma \Delta W_n + \Delta \tilde J_n \; = \; \Delta \Xi_n + \Delta \tilde J_n,
\end{cases}
\end{equation} 
where $\Delta X_n := X_{n+1} - X_{n} \, \in \Sigma_x$, $\hat \mu \in \R$ and $\hat \sigma > 0$. 
The term $\Delta \Xi_n := \hat \mu  \Delta t + \hat \sigma \Delta W_n$ takes values in $\{ -h_x, 0, h_x\}$\footnote{A common alternative is to consider a binomial discretization 
with $\Delta \Xi \in \{-h_x,h_x\}$, as in \cite{DaPaZa93}. }
and satisfies $\E\bigl[\Delta \Xi_n\bigr] = \hat \mu  \Delta t$ and $\E\bigl[(\Delta \Xi_n)^2\bigr] = \hat \sigma \Delta t$, at first order in $\Delta t$.  
The term $\Delta \tilde J_n$ is the discrete version of the compensated Poisson jump term, and satisfies $\E\bigl[\Delta \tilde J_n\bigr] = 0$ and 
$\E\bigl[(\Delta \tilde J_n)^2\bigr] = \tilde \sigma \Delta t$, at first order in $\Delta t$, with $\tilde \sigma > 0$. 
When the continuous time jump term is $\int_{\R} z \tilde N(dt,dz)$, 
the corresponding discrete version $\Delta \tilde J_n$ can assume all the values in $ \Sigma_x $.
If instead the integral has a truncation term $\epsilon$, i.e. $\int_{|z| \geq \epsilon} z \tilde N(dt,dz)$, we can define the subset 
$ \Sigma^{\epsilon}_x := \Sigma_x \setminus \{ -h_x, 0, h_x\}$, such that $\Delta \tilde J_n \in \Sigma^{\epsilon}_x$.

The Markov chain $\{X_n\}_{n\in \N}$ has the shape of a recombining multinomial tree, where each node has $\bar L$ branches.  
The number of nodes at time $n$ is $n(\bar L-1)+1$.
We derive the transition probabilities by an explicit discretization of the infinitesimal generator (see Appendix \ref{B2}).
Following \cite{Kushner}, the process $\{X_n\}_{n\in \N}$ has to satisfy the following two conditions in order to be admissible:
\begin{enumerate}
 \item the transition probabilities have the representation:
 \begin{equation}\label{local1}
  p^X \bigl(X_n,X_{n+1}\bigr) = \bigl(1-\lambda \Delta t \bigr) p^{D}\bigl(X_n,X_{n+1}\bigr) + \bigl( \lambda \Delta t \bigr) p^J\bigl(X_n,X_{n+1}\bigr)
 \end{equation}
  where $\lambda > 0$, and $p^{D}$, $p^J$ are respectively the diffusion and jump transition probabilities (see 
  Appendix \ref{B1}).
 \item (\emph{local consistency}) The moments of the discrete increments match those of the continuous increments, at first order in $\Delta t$: 
\begin{equation}\label{local2}
  \mathbb{E}_n \bigl[ \Delta X_n \bigr] = \mathbb{E}_t \bigl[ \Delta X_t \bigr], \; \quad  
  \mathbb{E}_n \bigl[ ( \Delta X_n )^2 \bigr] = \mathbb{E}_t \bigl[ ( \Delta X_t )^2 \bigr].  
\end{equation}
\end{enumerate}

The process $\{Y_n\}_{n\in \N}$ assumes values in $\Sigma_y$ and $\Delta L_n$, $\Delta M_n$ are non-negative multiples of $h_y$.
The two increments $\Delta L_n := L(t_{n}) - L(t_{n}^-)$ and $\Delta M_n := M(t_{n}) - M(t_{n}^-)$ can occur instantaneously at time $t_n$. 
They cannot assume values different from $0$ at the same time, 
and must satisfy the condition $Y_{n+1} = Y_n + \Delta Y_n \in \Sigma_y$\footnote{The values attainable by $\Delta L_n$ and $\Delta M_n$ 
depend on the current value of $Y_n \in \Sigma_y$. 
For instance, if $Y_n = -K_3 h_y$, then $\Delta L_n \in \{0, h_y, ...,(\bar M-1) h_y \}$ and $\Delta M_n \in \{0\}$.} for all $n$.

\subsection{Discrete dynamic programming algorithm}\label{algorithm_Sect}
We can formulate a discrete backward algorithm by applying the dynamic programming principle to (\ref{minimization}) 
on the discrete nodes of the chain $\{( Y_n,X_n )\}_n$:
\begin{align}\label{HJB3}
 & Q^{j}(t_n,Y_n,X_n) = \min  
 \; \biggl\{ \E_n \biggl[ Q \bigl( t_{n+1}, Y_n, X_n + \Delta X_n \bigr) \biggr], \\ \nonumber
 & \min_{\Delta L_n} \, \exp \biggl(\frac{\gamma}{\delta(t_n,T)} (1+\theta_b) e^{X_n} \Delta L_n \biggr) 
  \E_n \biggl[ Q^{j} \bigl( t_{n+1}, Y_n+\Delta L_n, X_n + \Delta X_n \bigr) \biggr], \\ \nonumber
 & \min_{\Delta M_n} \, \exp \biggl(\frac{-\gamma}{\delta(t_n,T)} (1-\theta_s) e^{X_n} \Delta M_n \biggr)
  \E_n \biggl[ Q^{j} \bigl( t_{n+1}, Y_n-\Delta M_n, X_n + \Delta X_n \bigr) \biggr]
 \biggr\}.
\end{align}
The variations of $\{Y_t\}_{t \in [t_0,T]}$ are instantaneous at $t_n$ for each $n$, 
while the process $\{X_t\}_{t \in [t_0,T]}$ changes in the interval $[t_n,t_{n+1}]$ according to its Lévy dynamics.
This feature suggests to introduce a numerical scheme based on two steps: an evolution step and a control step.

From now on we drop the superscript $j$ from $Q^j$. We introduce the discretization parameter $\rho = (\Delta t, h_x, h_y)$ and indicate the discretized value function by 
$Q^{\rho}$. For a fixed $\rho$, we adopt the common short notation $Q^n_{j,i} := Q^{\rho}(t_n,y_j,x_i)$. \\
We set the initial value $x_i = X_{n=0}$ for $i=0$. At time $n$, the index $i$ assumes values in 
$\{ -n K_1,-n K_1+1,...,n K_2-1, n K_2 \}$ and $j$ assumes values in $\{ -K_3, -K_3+1,..., K_4-1, K_4 \}$.   
\begin{scheme}\label{scheme}
Let us define a two steps numerical scheme $\mathbb{S}$ such that 
\begin{equation}\label{scheme1}
 \mathbb{S} \bigl( \rho, (t_n,y_j,x_i), Q^{\rho}(t_n,y_j,x_i), [Q^{\rho}]_{t_n,y_j,x_i} \bigr) = 0,
\end{equation}
where $[Q^{\rho}]_{t_n,y_j,x_i}$ indicates all the values of $Q^{\rho}$ not in $(t_n,y_j,x_i)$.  
\begin{align*}
 & \mbox{\textbf{step 1: }} \quad Q^n_{j,i} = \sum_{k = -K_1}^{K_2} p_k \; Q^{n+1}_{j,i+k} \quad  \mbox{for all} \quad j,i \\  
 & \mbox{\textbf{step 2: }} \quad \mathbb{S} = 
  Q^{n}_{j,i} - \min \biggl\{ Q^n_{j,i} , \, \min_{l} F(x_i,l,n) Q^n_{j+l,i}, \, \min_{m} G(x_i,m,n) Q^n_{j-m,i}  \biggr\} 
\end{align*}
such that $l\in \{0,...,K_4-j\}$ and $m\in \{0,...,K_3+j\}$ for each $j$. 
We defined $F(x_i,l,n) := e^{\bigl(\frac{\gamma}{\delta(t_n,T)} (1+\theta_b) e^{x_i} lh_y \bigr)}$ and 
$G(x_i,m,n) := e^{\bigl(- \frac{\gamma}{\delta(t_n,T)} (1-\theta_s) e^{x_i} mh_y \bigr)}$.
The coefficients $p_k$ satisfy $\sum_{k = -K_1}^{K_2} p_k = 1$ and $p_k \geq 0$ for all $-K_1 \leq k \leq K_2$. 
\end{scheme}
\begin{theorem}\label{theorem1}
 The Scheme [\ref{scheme}] (with $p_k$ defined in the Appendix, Equation (\ref{pK}) ) is monotone, stable and consistent. 
\end{theorem}
\begin{theorem}\label{theorem2}
 The solution $Q^{\rho}$ of (\ref{scheme1}) converges uniformly to the unique viscosity solution of (\ref{HJB2}).
\end{theorem}
\noindent
We prove Theorems [\ref{theorem1}] and [\ref{theorem2}] in the Appendix \ref{C}. 

\begin{algorithm}[H] 
\caption{Backward algorithm}
\label{algo}
 \algsetup{indent=1.5em}
 \begin{algorithmic}[1]
    \REQUIRE $r, (b,\sigma,\nu), X_0, K, T, \theta_b, \theta_s, \gamma, N, \bar L, \bar M. $
    \ENSURE $Q^j(t_0,y,X_0)$ for $j=0,w,b$
      \STATE Create the lattice for (\ref{log_sde_discr}) with appropriate discrete steps $\Delta t, h_y, h_x$.
      \STATE Create a vector of transition probabilities $p_k$ (as in \ref{pK} for instance).  
      \STATE Use (\ref{terminal_c}) or (\ref{terminal_w}) or (\ref{terminal_b}) to initialize a $\bar M \times \bigl( N(\bar L-1)+1 \bigl)$ grid for $Q^N_{j,i}$.  
      \FOR {n = N-1 to 0}
      \STATE $W_{j,i} = \sum_{k = -K_1}^{K_2} p_k \; Q^{n+1}_{j,i+k}$
      \STATE $Q^{n}_{j,i} = \min \biggl\{ W_{j,i} , \, \min_l F(x_i,l,n) W_{j+l,i}, \, \min_m G(x_i,m,n) W_{j-m,i}  \biggr\}$ 
      \ENDFOR
  \end{algorithmic}
\end{algorithm} 

\noindent
The computational complexity of the Algorithm [\ref{algo}] is 
$$\mathcal{O}\biggl( (N+1)\bigl[\frac{N(\bar L-1)}{2}+1 \bigr] \times \bar M \times \bar M \biggr).$$
The first factor comes from the loop over all the nodes of the tree i.e. $\sum_{n=0}^N n(\bar L-1)+1$. The second factor, $\bar M$, comes from the loop over all the values $y_j$, 
and the third factor, $\bar M$, comes from the minimum search. 

For a simple diffusion process the number of branches is fixed to $\bar L = 3$, but for processes with jumps it is proportional to $\sqrt N$.
The standard deviation of every Lévy process satisfying (\ref{fin_moment}) grows as the square root of time.
Therefore the size of a space step $h_x \propto \sqrt{\E[\Delta X^2]} \propto \sqrt{\Delta t} \propto \frac{1}{\sqrt{N}}$.
Let us consider for instance the integral term in Eq. (\ref{log_sde_Merton}) or (\ref{log_sde_inf_act}).
For computational reasons we have to reduce the region of integration to the bounded domain $[-B_1,B_2]$, with $B_1,B_2>0$ (see Appendix \ref{B2}). The number of branches to cover this
region is $\bar L = \frac{B_1+B_2}{h_x} \propto \sqrt{N}$. 

In order to have a more accurate result, it is better to choose $h_y \propto h_x$ and consequently $\bar M \propto N$. 
In this way, the number $h_y$ of shares to buy or sell is more sensitive to the resolution $h_x$ in the log-price tree.
Therefore assuming $\bar L \propto \sqrt{N}$ and $\bar M \propto N$, the total computational complexity is $\mathcal{O}(N^{4.5})$. 
For a fixed $\bar L$, the total complexity is reduced to $\mathcal{O}(N^{4})$.

\subsection{The Merton jump-diffusion model}

The first jump-diffusion model applied to finance is the \emph{Merton model}, presented in \cite{Me76}. 
In the paper, the author derives a semi-closed form solution for the price of a European call option. 
The Merton model describes the log-price evolution as a Lévy process $\{X_t\}_{t \in [t_0,T]}$ with a characteristic Lévy triplet $(b,\sigma,\nu)$
with $b \in \R$, $\sigma > 0$ and Lévy measure:
\begin{equation}\label{Merton_levy}
\nu(dz) = \frac{\lambda}{\xi \sqrt{2\pi}} e^{- \frac{(z-\alpha)^2}{2 \xi^2}} dz. 
\end{equation}
The process $\{X_t\}_{t \in [t_0,T]}$ can be represented as the superposition of a Brownian motion with drift and a pure jump process.
The number of jumps is represented by a Poisson process with intensity $\lambda>0$. The size of the jumps is 
normal distributed $\sim \mathcal{N}(\alpha, \xi^2)$. 
The dynamics of $\{X_t\}_{t \in [t_0,T]}$ is described by the SDE (\ref{log_sde_Merton}) with $ m = \lambda \bigl( e^{\alpha+\frac{\xi^2}{2}} -1 \bigr) $.
For any $f\in C^2(\R) \bigcap C_2(\R)$, the associated infinitesimal generator is:
\begin{align}\label{infinitesimal_M}
 \LL^{Mert} f(x) &= \; \biggl( \mu - \frac{1}{2}\sigma^2 - m \biggr) \frac{\partial f(x)}{\partial x}
+ \frac{1}{2} \sigma^2 \frac{\partial^2  f(x)}{\partial x^2} \\ \nonumber
&- \lambda f(x) + \int_{\R} f(x+z)\, \nu(dz).
\end{align}

\subsection{The Variance Gamma model} 

The \emph{Variance Gamma} (VG) process is a pure jump L\'evy process with infinite activity and no diffusion component. Applications of the VG process to financial 
modeling can be found for example in \cite{MaSe90} and \cite{MCC98}. 
Consider a Brownian motion with drift $Z_t = \theta t + \bar \sigma W_t$ and substitute the time variable $t$ with a gamma subordinator
$T_t \sim \Gamma(t,\kappa t)$. We obtain the VG process
$Z_{T_t} = \theta T_t + \bar \sigma W_{T_t}$,
that depends on three parameters:
$\theta$ is the drift of the Brownian motion,
$\bar \sigma$ is the volatility of the Brownian motion and
$\kappa$ is the variance of the Gamma process.
The VG L\'evy measure is:
\begin{equation}\label{VG_measure}
 \nu(dz) = \frac{e^{\frac{\theta z}{\bar \sigma^2}}}{\kappa|z|} \exp 
 \left( - \frac{\sqrt{\frac{2}{\kappa} + \frac{\theta^2}{\bar \sigma^2}}}{\bar \sigma} |z|\right) dz.
\end{equation}  
The L\'evy triplet is $\bigl( \int_{|z|<1} z \nu(dz), 0, \nu \bigr)$.
A VG process with an additional drift $c\in \R$, $X_t = ct + Z_t$, has the L\'evy triplet $\bigl( c + \int_{|z|<1} z \nu(dz), 0, \nu \bigr)$.  
The VG SDE can be obtained directly from Eq. (\ref{portfolio_dynamics2}):
\begin{align}\label{VG_sde}
  dX_t = \biggl( \mu -\omega + \theta \biggr) dt + \int_{\R} z \tilde N(dt,dz)
\end{align}
where we put $\sigma=0$, $ \theta = \int_{\R} z \nu(dz)$, $\omega = \int_{\R} \bigl( e^z -1 \bigr) \nu(dz)$ 
and $\mu = c + \omega$ by (\ref{mu}).

However, the jump process in Eq. (\ref{VG_sde}) has infinite activity, and its infinitesimal generator is not helpful for the Markov chain approximation!\\ 
We consider instead the approximated process (\ref{log_sde_inf_act}) with $\sigma=0$.
All the parameters are obtained through the VG L\'evy measure (\ref{VG_measure}):
\begin{equation}\label{log_sde_VG}
dX_t = \biggl( \mu - \frac{1}{2} \sigma_{\epsilon}^2 - \omega_{\epsilon} + \lambda_{\epsilon} \theta_{\epsilon}  \biggr) dt 
       + \sigma_{\epsilon} dW_t + \int_{|z|\geq \epsilon} z \tilde N(dt,dz)
\end{equation}
For any $f\in C^2(\R) \bigcap C_2(\R)$, the associated infinitesimal generator has a jump-diffusion form:
\begin{align}\label{VG_inf_gen}
\LL^{VG} f(x) \; =& \; \bigl( \mu-\frac{1}{2}\sigma_{\epsilon}^2 - w_{\epsilon} \bigr) \frac{\partial f}{\partial x} 
+ \frac{1}{2}\sigma_{\epsilon}^2 \frac{\partial^2 f}{\partial x^2} \\ \nonumber
&+ \int_{|z| \geq \epsilon} f(x+z) \nu(dz) - \lambda_{\epsilon} f(x).
\end{align}

\section{Numerical results}\label{numerical}

In this section we implement the Algorithm [\ref{algo}] described in Section \ref{algorithm_Sect} and calculate the prices of European call options for the writer and the buyer.
The prices are computed under the assumption that the stock log-price follows three different L\'evy processes: a Brownian motion, a Merton jump-diffusion and a Variance Gamma, with
parameters in Table \ref{tab:parameters}.
\begin{table}[ht]
\centering
 \begin{tabular}[t]{*{11}l}
 \toprule
  \multicolumn{5}{l}{\textbf{Details}} & \multicolumn{6}{l}{\textbf{Diffusion parameters}} \\
  \midrule
  $K$ & $T$ & $r$ & & & $\mu$ & $\sigma$ & $\gamma$ \\
  15 & 1 & 0.1 & & & 0.1 & 0.25 & 0.001 \\
  \toprule
  \multicolumn{5}{l}{} & \multicolumn{6}{l}{\textbf{Merton parameters}} \\
  \midrule
  & & & & & $\mu$ & $\sigma$ & $\alpha$ &$\xi$ & $\lambda$ & $\gamma$\\
  & & & & & 0.1 & 0.25 & 0 & 0.5 & 0.8 & 0.04\\
  \toprule
  \multicolumn{5}{l}{} & \multicolumn{6}{l}{\textbf{VG parameters}} \\
  \midrule
  & & & & & $\mu$ & $\theta$ & $\bar \sigma$ &$\kappa$ & $\gamma$ \\
  & & & & & 0.1 & -0.1 & 0.2 & 0.1 & 0.05 \\
\bottomrule
\end{tabular}
  \caption{Option details and parameters for diffusion, Merton and VG processes.}
  \label{tab:parameters}
\end{table}

\begin{table}[ht]
\centering
\begin{tabular}[t]{llllll}
\toprule
  \multicolumn{2}{c}{\textbf{Diffusion price}} &  \multicolumn{2}{c}{\textbf{Merton price}} &  \multicolumn{2}{c}{\textbf{VG price}} \\
\midrule
Closed formula & PDE & Closed formula & PIDE & Closed formula & PIDE\\
2.2463 & 2.2463 & 3.4776 & 3.4775 & 1.9870 & 1.9871\\
\bottomrule
\end{tabular}
\caption{At the money prices with $S_0=K=15$ with parameters in Table \ref{tab:parameters}.}
\label{tab:ATM_price}
\end{table}%

\begin{figure}[t!]
 \begin{minipage}[b]{0.5\linewidth}
   \centering
   \includegraphics[width=\linewidth]{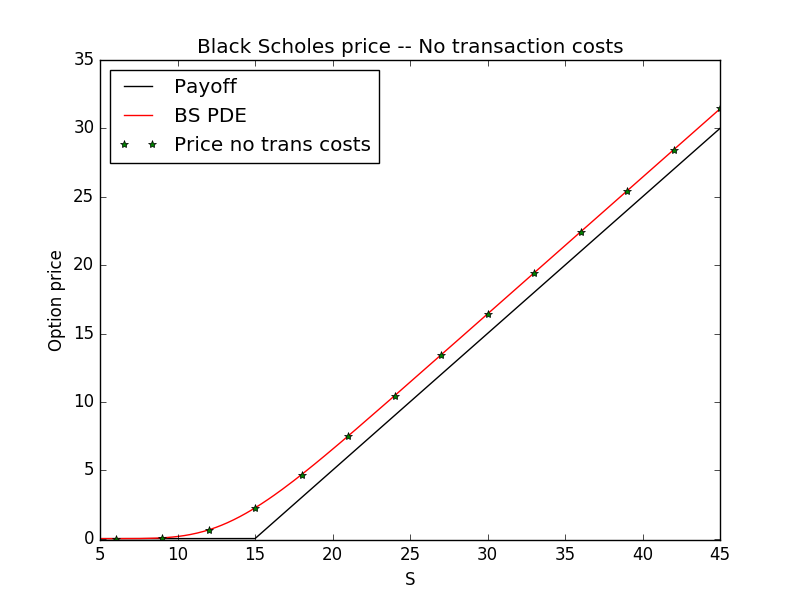}
 \end{minipage}
 \ \hspace{2mm} \hspace{3mm} \
 \begin{minipage}[b]{0.5\linewidth}
  \centering
   \includegraphics[width=\linewidth]{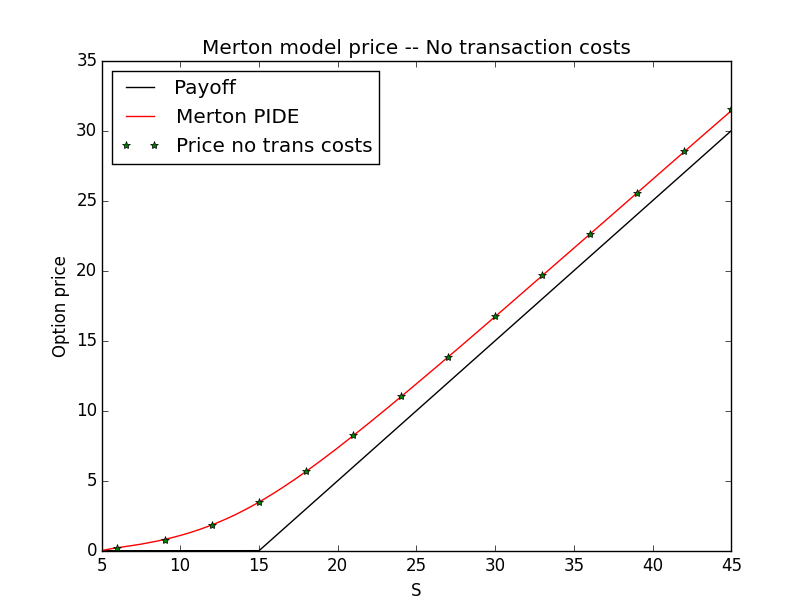}
 \end{minipage}
  \ \hspace{2mm} \hspace{3mm} \
  \begin{minipage}[b]{\linewidth}
  \centering
   \includegraphics[width=0.51\linewidth]{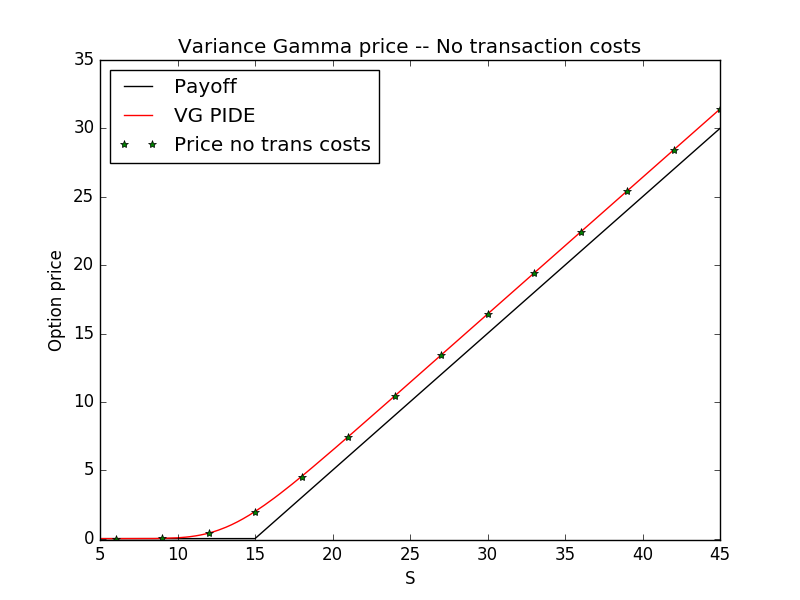}
   \caption{Writer prices with zero transaction costs for diffusion (top-left), Merton (top-right) and VG (bottom) process. Parameters are in Tab. \ref{tab:parameters}.}
   \label{Fig1}
 \end{minipage}
\end{figure}
For comparisons, we compute also the option prices using the standard 
\emph{martingale pricing theory} (see Appendix \ref{A}). In the Table \ref{tab:ATM_price} we show the \emph{at the money} values obtained with the closed formula 
and by solving the respective PIDE. 
The closed formula for the diffusion process is the well known \cite{BS73} formula. To compute the Merton price we use the semi-closed formula derived in \cite{Me76},
and for the VG price we used the semi-closed formula derived in \cite{MCC98}. The PIDE prices are obtained by solving the Eq. (\ref{PIDE}) with generators (\ref{infinitesimal_BS}),
(\ref{infinitesimal_M}) and (\ref{VG_inf_gen}) (details in Appendix \ref{A}).
Of course, the parameter $\mu$ has not been used to compute the prices in Tab. \ref{tab:ATM_price}. 
We set the drift term $\mu$ equal to the risk free interest rate $r$, following the common rule of the standard no-arbitrage theory.
This choice is also discussed in \cite{HoNe89},     

In the following analysis, we consider the PIDE prices as our benchmarks for comparisons.   
In all the computations we use equal transaction costs for buying and selling, $\theta_b = \theta_s$.

In Fig. \ref{Fig1} we show that the proposed model prices replicate the PIDE prices for zero transaction costs and small values of $\gamma$. 
The values of $\gamma$ in Table \ref{tab:parameters},
are chosen very small\footnote{ In chapter 5 of \cite{GK99} are presented some common values for the risk aversion coefficient: $\gamma=0.3$, $\gamma=0.2$ and $\gamma=0.1$ 
for high, medium and low level of risk aversion respectively.} for this purpose. 
An intuitive argument to justify this choice is that for $\gamma \to 0$, the utility function 
can be approximated by a linear utility $\mathcal{U}(w) = 1 - e^{-\gamma w} \approx \gamma w$ and the investor can be considered risk neutral. 
A rigorous argument can be found in \cite{BaSo98}, where the authors use asymptotic analysis for small values of $\theta_b$, $\theta_s$ and $\gamma$ to derive 
a nonlinear PDE for the option price. For zero transaction costs this equation corresponds to the Black-Scholes PDE. 
Their argument can be extended also to PIDEs.
\begin{figure}[t!]
 \begin{minipage}[b]{0.5\linewidth}
   \centering
   \includegraphics[width=\linewidth]{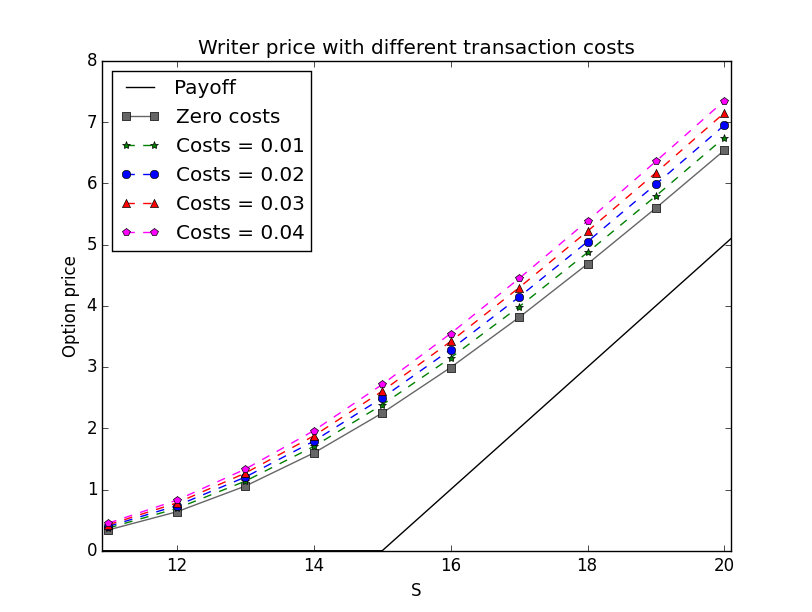}
 \end{minipage}
 \ \hspace{2mm} \hspace{3mm} \
 \begin{minipage}[b]{0.5\linewidth}
  \centering
   \includegraphics[width=\linewidth]{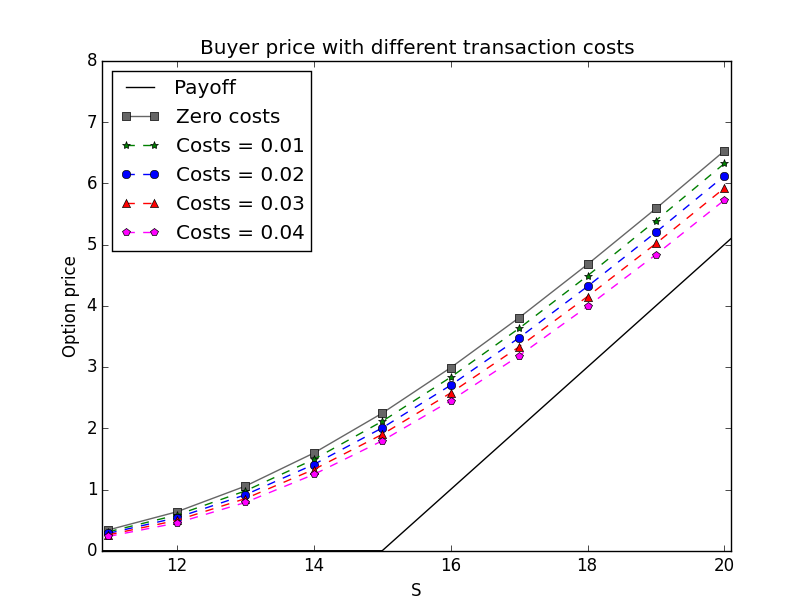}
 \end{minipage}
    \caption{Writer and buyer prices for different levels of transaction costs. The continuous line is the solution of the Black-Scholes PDE.}
   \label{Fig2}
\end{figure}

\begin{table}[ht]
\centering
 \begin{tabular}{*{11}l}
 \toprule
  \multicolumn{5}{c}{\textbf{Convergence table}} \\
  \midrule
  $N = \bar M$ & $\gamma = 0.0001$ & $\gamma = 0.001$ & $\gamma = 0.01$ & Execution time \\
  \midrule
    50   & 2.241214 & 2.241764 & 2.247311 & 0.01 $\pm$ 0.004\\
    100  & 2.249142 & 2.249506 & 2.253159 & 0.02 $\pm$ 0.005\\
    200  & 2.245422 & 2.245676 & 2.248216 & 0.11 $\pm$ 0.02\\
    400  & 2.246784 & 2.246959 & 2.248717 & 0.85 $\pm$ 0.04 \\
    800  & 2.246288 & 2.246271 & 2.247635 & 8.63 $\pm$ 0.1 \\
    1600 & 2.246576 & 2.246662 & 2.247515 & 82.44 $\pm$ 2.71\\
    3200 & 2.246412 & 2.246471 & 2.247068 & 910.8 $\pm$ 10.5\\
    3500 & 2.246366 & 2.246423 & 2.246993 & 1291.3 $\pm$ 13\\
  \bottomrule
  \end{tabular}
  \caption{Convergence table for ATM diffusion prices with zero transaction costs.}
  \label{tab:convergence}
\end{table}

\subsection{Diffusion results}

In the Figure \ref{Fig2} we show the diffusion writer and buyer prices with different
transaction costs.  
We can see that a higher transaction cost corresponds to a higher writer price, while a lower transaction cost corresponds to a lower buyer price.
In fact, the writer and buyer prices are respectively increasing and decreasing functions of the transaction cost, as already verified in \cite{ClHo97}.
The prices in Figure \ref{Fig2}, are calculated with $N=1500$ time steps and $\bar M = N$. 
In the Table \ref{tab:convergence} we show ATM option prices for different values of $N$, with $\theta_s = \theta_b = 0$ and different risk aversion coefficients.
For $\gamma=0.0001$ and $N=3500$ the price is identical, up to the fourth decimal digit, to the original Black-Scholes price in Table \ref{tab:ATM_price}. 
Using the values in Table \ref{tab:convergence} it is possible to estimate the \emph{rate of convergence}.
We also present the execution times, from which we can estimate the asymptotic \emph{time complexity} of the algorithm. 
In Section (\ref{algorithm_Sect}) we stated that the computational complexity of the Algorithm [\ref{algo}] is $\mathcal{O}(N^{4})$. From the Table \ref{tab:convergence}, 
we obtain the exponent $\frac{\log(1291.3/910.8)}{\log(3500/3200)} = 3.9$, which is very close to the theoretical complexity value.  
The algorithm is written in Matlab using vectorized operations, and runs on an Intel i7 (7th Gen) with Linux.

\subsection{Merton results}

\begin{figure}[t!]
 \begin{minipage}[b]{0.5\linewidth}
   \centering
   \includegraphics[width=\linewidth]{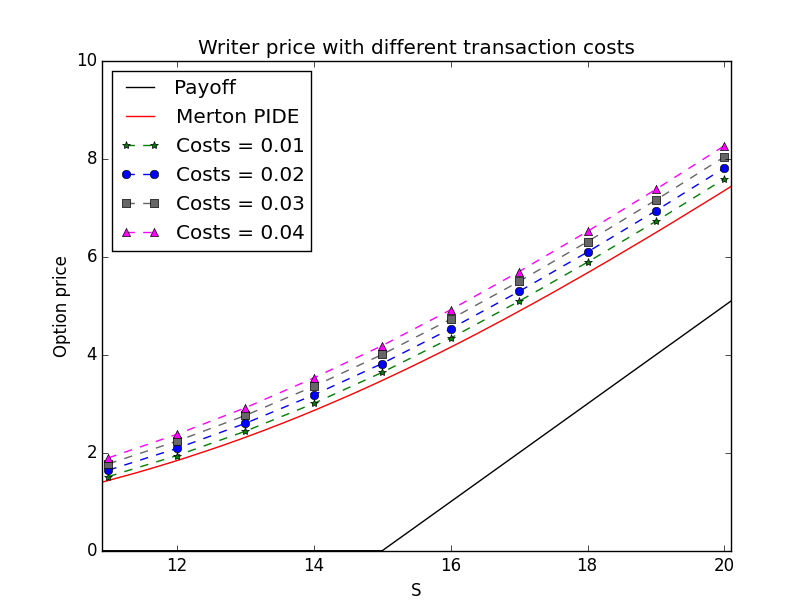}
 \end{minipage}
 \ \hspace{2mm} \hspace{3mm} \
 \begin{minipage}[b]{0.5\linewidth}
  \centering
   \includegraphics[width=\linewidth]{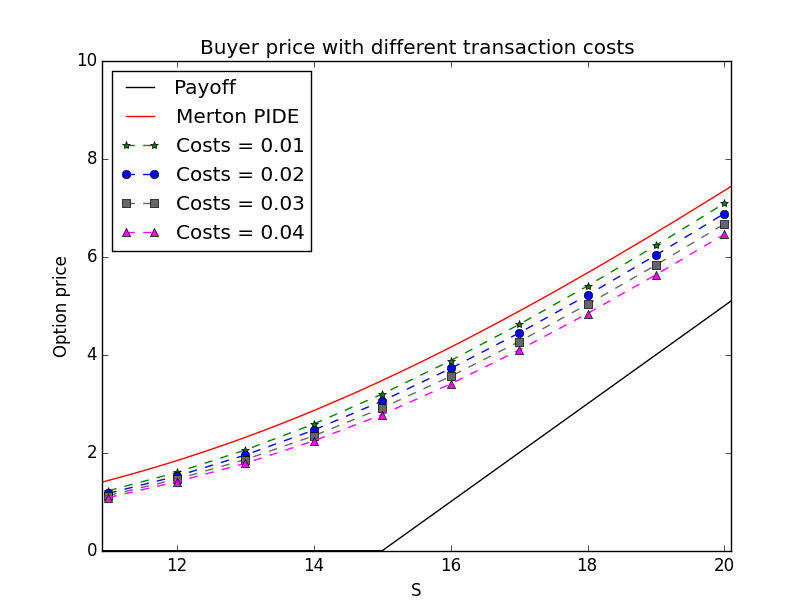}
 \end{minipage}
 \caption{Writer and buyer prices for different transaction costs. The continuous line is the solution of the Merton PIDE.}
   \label{Fig3}
\end{figure}
\begin{figure}[t!]
 \begin{minipage}[b]{0.5\linewidth}
   \centering
   \includegraphics[width=\linewidth]{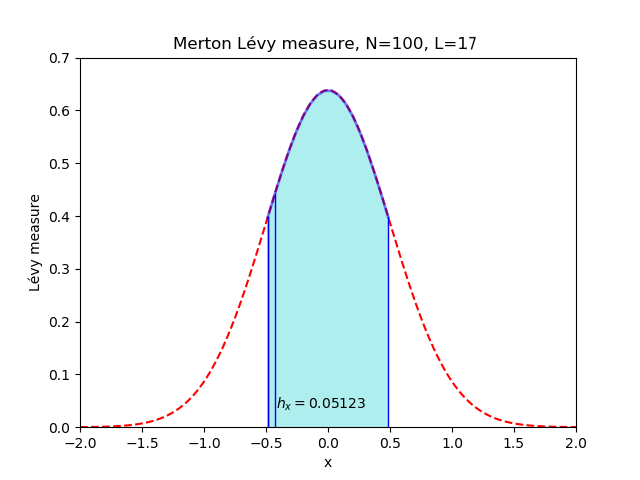}
   \caption{Merton L\'evy measure computed using parameters in Tab. \ref{tab:parameters}, $N=100$ and $\bar L=17$. 
   The domain $[-B_1,B_2] = [-\sqrt{\lambda} \xi,\sqrt{\lambda} \xi]$ has length $\bar L h_x \approx 2 \sqrt{\lambda} \xi$.}
   \label{Fig13} 
 \end{minipage}
 \ \hspace{2mm} \hspace{3mm} \
 \begin{minipage}[b]{0.5\linewidth}
  \centering
   \includegraphics[width=\linewidth]{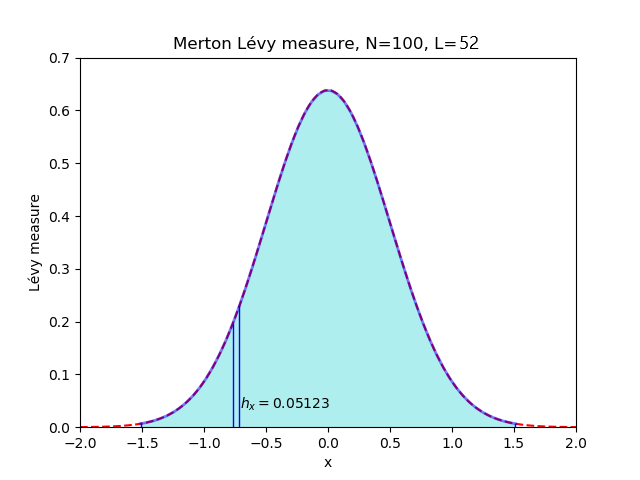}
   \caption{Merton L\'evy measure computed using parameters in Tab. \ref{tab:parameters}, $N=100$ and $\bar L=52$. 
   The domain $[-B_1,B_2] = [-3\sqrt{\lambda}\xi,3\sqrt{\lambda}\xi]$ has length $\bar L h_x \approx 6 \sqrt{\lambda} \xi$.}
   \label{Fig14}
 \end{minipage}
\end{figure}

In the Figure \ref{Fig3} we show the writer and buyer prices for the Merton process, with parameters in Tab. \ref{tab:parameters}.
An interesting feature of the multinomial tree construction for jump-diffusion processes is that $\bar L \propto \sqrt{N}$.
The integral domain is restricted to the bounded domain $[-B_1,B_2]$ with length $B_1+B_2 = \bar L \, h_x$. 
We choose the size of a space step $h_x = \sqrt{\E[\Delta X^2]} = \sigma_X \sqrt{\Delta t}$ and $\sigma_X^2 = \sigma^2 + \tilde \sigma_J^2$ 
with $ \tilde \sigma_J^2 = \int_{-B_1}^{B_2} z^2 \nu(dz)$.
However, the size of the Poisson jumps does not scale with $\Delta t$. So the number $\bar L$ has to be chosen big enough in order 
to have $L\, h_x \geq B_1+B_2$. 
In general, for a fixed $h_x$, the interval $[-B_1,B_2]$ should be chosen as big as possible. In practice, the choice of the truncation depends on the shape of the L\'evy measure.  
The Figures \ref{Fig13}, \ref{Fig14} show two examples with $[-B_1,B_2] = [\sqrt{\lambda} \xi,\sqrt{\lambda} \xi]$ and $[-B_1,B_2] = [-3\sqrt{\lambda} \xi,3\sqrt{\lambda} \xi]$ 
respectively. These choices correspond to 
$\bar L = 17$ and $\bar L = 52$.\\
For the Merton L\'evy measure (a scaled Normal distribution), a good choice is $[-B_1,B_2] = [-3\sigma_J,3\sigma_J]$, where  
$\sigma_J^2 = \int_{\R} z^2 \nu(dz) = \lambda(\alpha^2 + \xi^2)$ is the variance of the jump component of the Merton process.
The length of the interval is thus $\bar L \, h_x = 6\sigma_J$. 
It is well known that the integral over this region is about the $99.74\%$ of the total area.      
Using this interval and the parameters in Tab. \ref{tab:parameters} we obtain the relation $\bar L \geq 5.86 \sqrt{N}$.
In the calculation of the Merton prices in Fig. \ref{Fig3}, we used a discretization with $N=\bar M =100$, and $\bar L = 81$, 
with a good balance between small computational time and small price error. 
\begin{table}[ht]
\centering
 \begin{tabular}{*{6}l}
 \toprule
  \multicolumn{6}{c}{\textbf{Truncation error table}} \\
  \midrule
  $N$ & $\bar L=51$ &   $\bar L=71$   & $\bar L=91$ & $\bar L=101$ & $\bar L=111$ \\
  \midrule
    50   & 3.481318 & 3.481616 & 3.481617 & 3.481617 & 3.481617  \\
    100  & 3.468774 & 3.478806 & 3.479141 & 3.479146 & 3.479146  \\
    150  & 3.439090 & 3.474403 & 3.477574 & 3.477714 & 3.477742  \\
    200  & 3.399442 & 3.466338 & 3.476439 & 3.477234 & 3.477457 \\
  \bottomrule
  \end{tabular}
  \caption{Truncation error for ATM Merton prices with zero transaction costs.}
  \label{tab:convergence2}
\end{table}

The convergence Tab. \ref{tab:convergence2} shows different Merton prices for different values of $N$ and $\bar L$.
Looking at the table from left to right, for each fixed $N$ it is possible to note how the truncation error decreases when $\bar L$ increases.
\begin{table}[ht]
\centering
 \begin{tabular}{llll}
 \toprule
  \multicolumn{4}{c}{\textbf{Convergence table}} \\
  \midrule
  $N = \bar M$ & $\bar L$ & Price & Execution time \\
  \midrule
    50  & 61  & 3.481600 & 2.20 $\pm$ 0.08 \\
    75  & 75  & 3.479980 & 15.15 $\pm$ 0.07 \\
    100 & 91  & 3.479141 & 63.04 $\pm$ 0.49 \\
    125 & 97  & 3.478254 & 148.4 $\pm$ 1.16 \\
    150 & 105 & 3.477731 & 315.3 $\pm$ 5.58 \\
    175 & 113 & 3.477610 & 585.7 $\pm$ 10.57 \\ 
    200 & 121 & 3.477513 & 1106.0 $\pm$ 12.2 \\
  \bottomrule
  \end{tabular}
  \caption{Convergence table for ATM Merton prices with zero transaction costs.}
  \label{tab:convergence3}
\end{table}

In Table \ref{tab:convergence3} we show several prices with increasing values of $N$ and $\bar L$. We choose $\bar L$ big enough, such that the truncation error can be ignored.
Given the high computational complexity of the algorithm, 
it is difficult to present prices with bigger values of $N$,$\bar L$. For larger values of $N$ and smaller $\gamma$, we expect a convergence to the Merton price in Tab. 
\ref{tab:ATM_price}. 
The computational complexity in this case is expected to be $\mathcal{O}(N^{4.5})$. From the Table \ref{tab:convergence3} we get 
the exponent equal to $\frac{\log(1106.0/585.7)}{\log(200/175)} = 4.76$. This result is not so different from the theoretical value.

\subsection{VG results}

\begin{figure}[t!]
 \begin{minipage}[b]{0.5\linewidth}
   \centering
   \includegraphics[width=\linewidth]{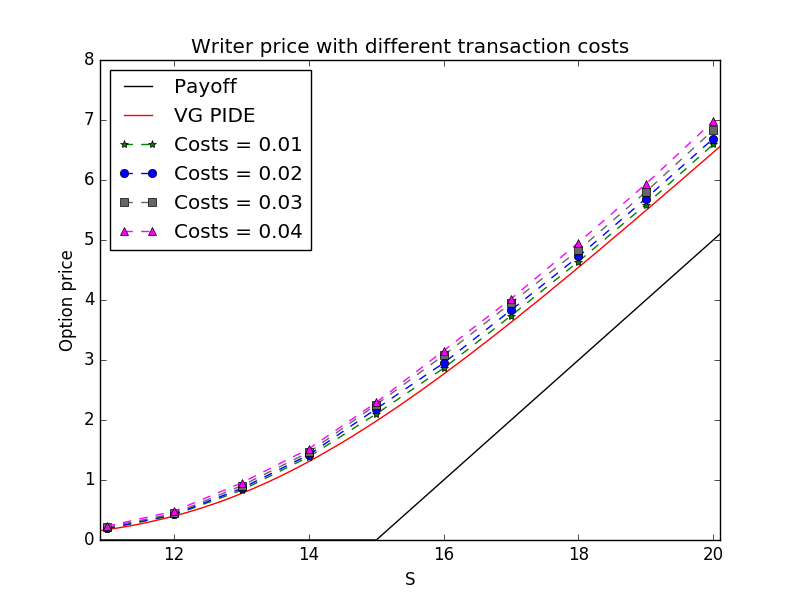}
 \end{minipage}
 \ \hspace{2mm} \hspace{3mm} \
 \begin{minipage}[b]{0.5\linewidth}
   \includegraphics[width=\linewidth]{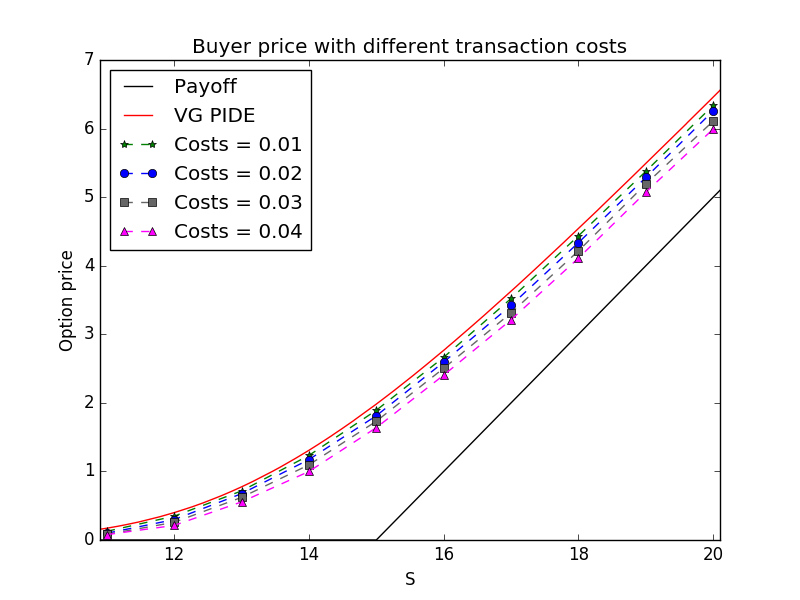}
 \end{minipage}
 \caption{Writer and buyer prices for different transaction costs. The continuous line is the solution of the VG PIDE.}
 \label{Fig8}
\end{figure}  

\begin{figure}[t!]
 \begin{minipage}[b]{0.5\linewidth}
   \centering
   \includegraphics[width=\linewidth]{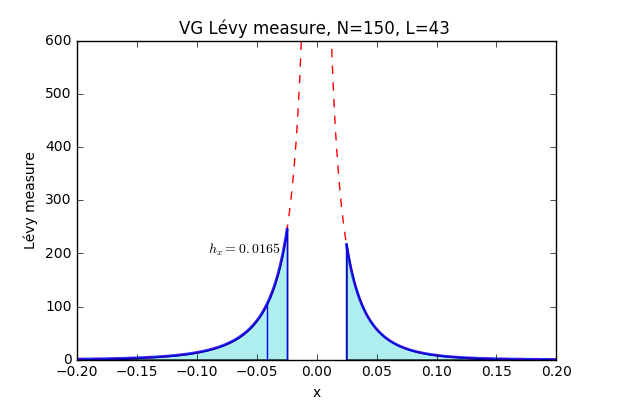}
   \caption{VG L\'evy measure computed using parameters in Tab. \ref{tab:parameters}, $N=150$ and $\bar L=43$. 
   The domain $[-B_1,-\epsilon]\bigcup [\epsilon,B_2]$ has length $(\bar L-3) h_x$. The highlighted area is $99.9\%$ of the total area.}
   \label{Fig15} 
 \end{minipage}
 \ \hspace{2mm} \hspace{3mm} \
 \begin{minipage}[b]{0.5\linewidth}
  \centering
   \includegraphics[width=\linewidth]{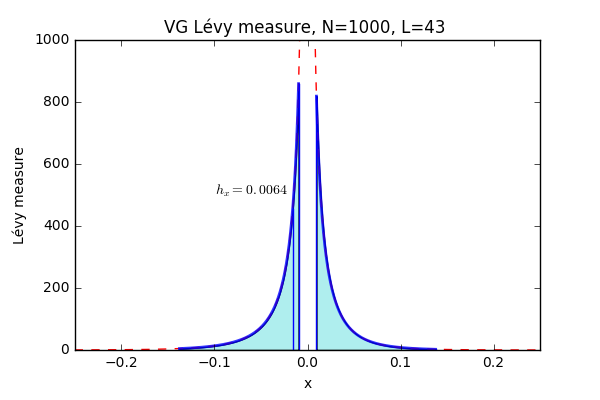}
   \caption{VG L\'evy measure computed using parameters in Tab. \ref{tab:parameters}, $N=1000$ and $\bar L=43$. 
   The domain $[-B_1,-\epsilon]\bigcup [\epsilon,B_2]$ has length $(\bar L-3) h_x$. The highlighted area is $98.9\%$ of the total area.}
   \label{Fig16}
 \end{minipage}
\end{figure}

In the Figure \ref{Fig8} we show how the writer and buyer prices for the VG process change for several level of transaction costs (parameters in Tab. \ref{tab:parameters}). 
In these computations we used $N=\bar M = 150$ and $\bar L = 43$, such that the program can run in a reasonable amount of time. 
The integration region in (\ref{VG_inf_gen}) is restricted to $[-B_1,-\epsilon]\bigcup [\epsilon,B_2]$ with $\epsilon=1.5h_x$. The choice of $B_1$ and $B_2$ depends on the shape of the 
L\'evy measure. 
In Fig. \ref{Fig15} and \ref{Fig16} we show two examples for the VG L\'evy measure with $N=150$, $h_x=0.0165$ and $N=1000$, $h_x=0.0064$. 
The two L\'evy measures are normalized, such that the integral on the region $[-\infty,-\epsilon]\bigcup [\epsilon,+\infty]$ is equal to one. The 
area underlying the functions on $[-B_1,-\epsilon]\bigcup [\epsilon,B_2]$ is highlighted for clarity.
For $\bar L = 43$, we can see that in both cases it is possible to cover a very high percentage of the initial unrestricted region. We can conclude that, unlike the Merton
measure, we do not need a big truncation interval.
Given the space step $h_X = \sigma_X\sqrt{\Delta t}$, with  
$ \sigma_X^2 = \hat \sigma_{J}^2 + \sigma_{\epsilon}^2 $ and $\hat \sigma_J^2 = \int_{[-B_1,-\epsilon]\bigcup [\epsilon,B_2]} z^2 \nu(dz)$, 
it is enough to consider a region at least as big as the standard deviation  
of the unrestricted jump process\footnote{For small values of $\epsilon$ the value of $\sigma_{\epsilon}^2$ is negligible. In this case it is possible to replace $\sigma_J^2$ 
by the variance of the VG process $\sigma_{VG}^2 = \bar \sigma^2 + \theta^2 \kappa$.}
i.e. $h_X \bar L \geq \sigma_J$, where $\sigma_J^2 = \int_{[-\infty,-\epsilon]\bigcup [\epsilon,\infty]} z^2 \nu(dz)$. 
Putting all together, the relation becomes $\bar L\geq \frac{\sigma_J}{\sigma_X} \sqrt{N}$, and replacing the values $\sigma_X = 0.2024$ and $\sigma_J=0.1916$ 
we get $\bar L \geq 0.94 \sqrt{N}$.
\begin{table}[ht]
\centering
 \begin{tabular}{llll}
 \toprule
  \multicolumn{4}{c}{\textbf{Convergence table}} \\
  \midrule
  $N = \bar M$ & $\lambda_{\epsilon}$ & Price & Execution time \\
  \midrule
    50  & 4.73  & 1.910934 & 3.63 $\pm$ 0.16 \\
    100 & 7.82  & 1.957806 & 26.54 $\pm$ 0.26 \\
    150 & 10.01 & 1.982078 & 82.51 $\pm$ 0.20 \\
    200 & 11.73 & 1.996180 & 185.2 $\pm$ 0.81 \\
    250 & 13.14 & 2.004719 & 350.3 $\pm$ 4.5 \\
    300 & 14.35 & 2.008536 & 654.2 $\pm$ 7.3\\ 
    350 & 15.40 & 2.009436 & 1236 $\pm$ 12 \\
  \bottomrule
  \end{tabular}
  \caption{Convergence of ATM VG prices with $\bar L =43$, zero transaction costs.}
  \label{tab:convergence4}
\end{table}

In the Table \ref{tab:convergence4} we present several option prices computed with different values of $N$, but with fixed $\bar L$.
For the VG process it is more difficult to analyze the convergence results.
This is due to the approximation (\ref{log_sde_inf_act}) introduced to replace the infinite activity jump component by a Brownian motion. All the parameters in
(\ref{sig_eps}) depend on $\epsilon$, and consequently on $N$.
Within our discretization ($N=150$), we have $\sigma_{\epsilon} = 0.0654$, $\lambda_{\epsilon} = 10.01$. 
With the parameters under consideration, we obtain an ATM price for zero transaction costs of 1.9821, which is quite close to the PIDE price in Tab. \ref{tab:ATM_price}. 


The convergence rate of the VG PIDE is quite low and this is reflected in our algorithm. We refer to \cite{CoVo05b} for a detailed error analysis.
In order to solve the PIDE (using an implicit-explicit scheme) we constructed
a grid with 13000 space steps of size $\delta x = 0.0004$ and 7000 time steps, and obtained the price in Tab \ref{tab:ATM_price}, 
with an approximated activity $\lambda_{\epsilon} = 75$.
Consequently, we expect to have good convergence results in our algorithm when $N\sim 10^4$.
All the presented prices (Figures \ref{Fig8}) have thus a truncation error, which is adjusted by an accurate choice of the value of $\gamma$. 

From Tab. \ref{tab:convergence4} we can estimate the time complexity of this algorithm. The exponent is $\frac{\log(1236.0/654.2)}{\log(350/300)} = 4.12$, 
indeed very close to the theoretical $\mathcal{O}(N^4)$.

\subsection{Properties of the model}\label{properties_model}

\begin{table}[ht]
  \centering
 \begin{tabular}{llllll}
\toprule
 & cost = 0 & cost = 0.01 & cost = 0.02 & cost = 0.03 & cost = 0.04  \\
\midrule
\textbf{Merton} & 3.4771 & 3.6400 & 3.8212 & 4.0054 & 4.1864 \\
\textbf{VG} & 1.9821 & 2.0921 & 2.1870 & 2.2568 & 2.3131 \\
\bottomrule
\end{tabular}
  \caption{Merton and VG writer prices for different transaction costs, with parameters as in Tab. (\ref{tab:parameters}). }
  \label{tab:costs}
\end{table}

In this section we want to analyze the properties of the model and how the option price depends on the level of transaction costs $\theta_b$, $\theta_s$, 
and on the risk aversion parameter $\gamma$. 
In this numerical experiment, we use the Merton model with parameters of Tab. \ref{tab:parameters}.
In Tab. \ref{tab:costs} we show the writer ATM option values for different transaction costs.  
\begin{figure}[t!]
 \begin{minipage}[b]{0.5\linewidth}
   \centering
   \includegraphics[width=\linewidth]{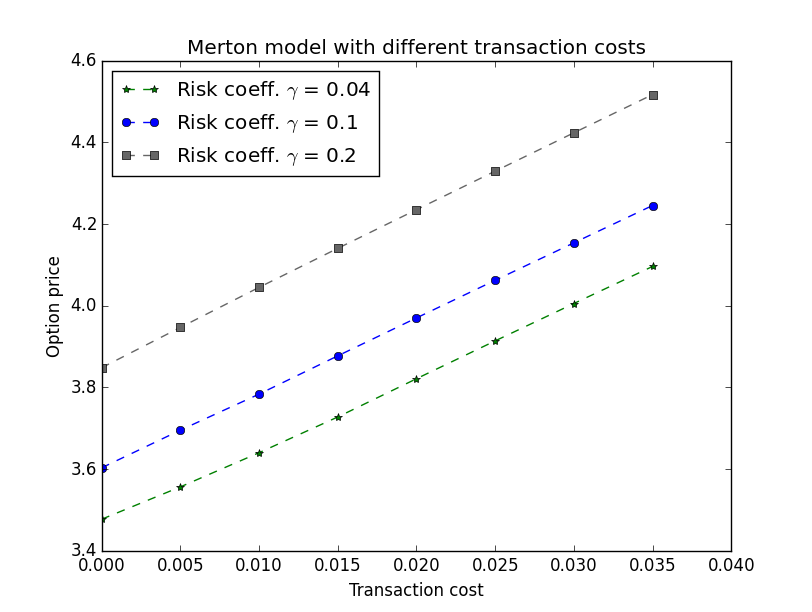}
   \caption{Merton option prices for the writer as function of the transaction cost, with different values of $\gamma$.}
   \label{Fig10} 
 \end{minipage}
 \ \hspace{2mm} \hspace{3mm} \
 \begin{minipage}[b]{0.5\linewidth}
  \centering
   \includegraphics[width=\linewidth]{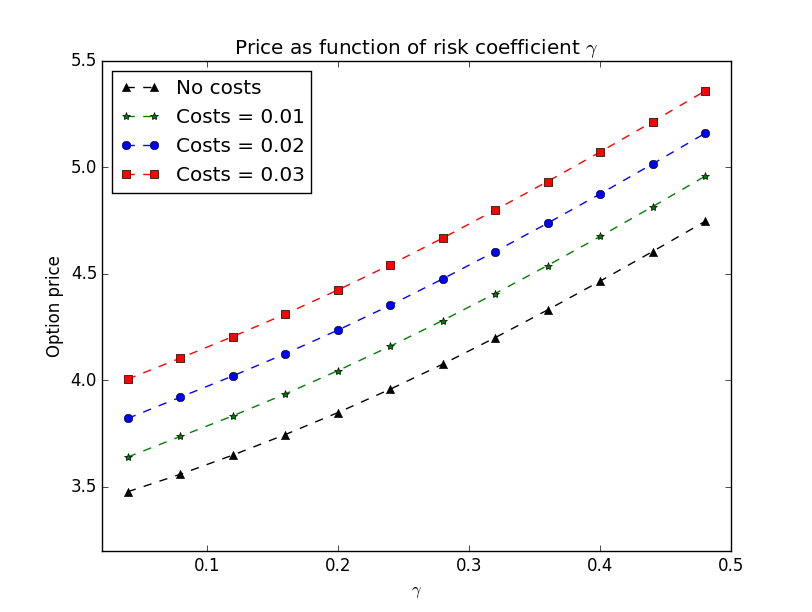}
   \caption{Merton option prices for the writer as function of $\gamma$, with different values of transaction costs.}
   \label{Fig11}
 \end{minipage}
\end{figure}

In Fig. \ref{Fig10} we can see better how the writer price is affected by the change of the transaction costs. 
The picture shows prices for different values of risk coefficient.
The risk profile of the investor also plays an important role. As already shown in \cite{HoNe89}, the writer price is an
increasing function of the risk aversion coefficient. Figure \ref{Fig11} confirms their results.


\section{Conclusions}\label{conclusions}

We presented a model for pricing options in presence of proportional transaction costs.
This is an extension of the model first introduced by \cite{HoNe89} and then formalized by \cite{DaPaZa93}. 
The main difference between this work and the previous works in the literature is that we considered a stock
dynamics that follows a general exponential Lévy process.
In the present paper, we do not consider the case of bankruptcy in the numerical computations. 
The general Eq. (\ref{HJB1}) is a complicated equation with three state variables and a time variable, 
so we opted to simplify the problem by reducing the number of variables, in order to obtain the simpler HJB Eq. (\ref{HJB2}).
The resulting optimization problem has been solved with the Markov chain approximation method. 
We proposed a monotone, stable and consistent numerical scheme and proved that its solution converges to the viscosity solution of the original
HJB equation.
Numerical results are obtained for diffusion, Merton jump-diffusion and Variance Gamma processes, although any Lévy process satisfying
the conditions (\ref{fin_moment}) can be used. The transition probabilities in the Markov chain approximation are obtained by explicit finite difference discretization of the 
infinitesimal generator of the process. 
The Brownian motion and the Merton process can be discretized directly, 
while the VG process needs to be approximated to remove the infinite activity jump component.
Due to this approximation, the Algorithm [\ref{algo}] has lower performances when applied to the VG dynamics.
Using numerical experiments, we confirmed some features of the model such as the relations of the price with the transaction costs and the risk aversion.

An interesting direction for future improvements, can be the development of a more efficient numerical method for solving the HJB Equation (\ref{HJB2}).
There are several approaches in the literature for solving variational inequalities, such as the policy iteration method of \cite{FoHu12a}, or the penalty method of 
\cite{FoHu12b}, \cite{Song14}.
We argue that a finite difference method based on the implicit/explicit scheme, 
with the possible help of the Fast-Fourier-Transform for evaluating the integral term, (as in \cite{AnAn00} for instance) 
can increase the efficiency of the numerical method.
Also, using a non-uniform grid as in \cite{Haentjens13}, can help to improve the efficiency 
and reduce the computational cost for both the differential and the integral part.

\section*{Acknowledgements}
Our sincere thanks go to the Department of Mathematics of ISEG and CEMAPRE, University of Lisbon, 
\url{http://cemapre.iseg.ulisboa.pt/}.
This research was supported by the European Union in the FP7-PEOPLE-2012-ITN project STRIKE - 
Novel Methods in Computational Finance (304617), \url{http://www.itn-strike.eu/}, and by CEMAPRE
MULTI/00491, financed by FCT/MEC through Portuguese national funds.
We wish also to acknowledge all the members of the STRIKE network.

\appendix
\numberwithin{equation}{section}

\section{Martingale option pricing}\label{A}
Under a risk neutral measure $\Q$, the dynamics of the stock price is described by the \emph{exponential Lévy model}:
 $S_t = S_0 e^{X_t} = S_0 e^{rt + L_t}$
where $r$ is the risk free interest rate, and $\{L_t\}_{t\in[t_0,T]}$ is a Lévy process with Lévy triplet $(b,\sigma,\nu)$. 
Under $\Q$ the discounted price is a $\Q$-martingale:
 $ \E^{\Q} \bigl[ e^{-rt} S_t \bigr| S_0 \bigr] =  \E^{\Q} \bigl[ S_0e^{L_t} \bigr| S_0 \bigr] = S_0 $,
such that $\E^{\Q}[ e^{L_t} | L_0=0] = 1 $. 
This implies the following condition:
\begin{equation}\label{martingale_b}
 b = -\frac{1}{2} \sigma^2  - \int_{\R} \bigl( e^z-1 -z\mathbbm{1}_{\{ |z|<1 \}} \bigr) \nu(dz) .
\end{equation}
Let $C(t,x) \in C^{1,2}\bigl( [t_0,T] \times \R \bigr) \bigcap C_2 \bigl( [t_0,T] \times \R \bigr)$ be the value of a European call option. 
It satisfies the following partial integro-differential equation (PIDE):
\begin{equation}\label{PIDE}
  \frac{\partial C(t,x)}{\partial t} + \LL^{X} C(t,x) -r C(t,x) = 0   
\end{equation}
where $\LL^{X}$ is the infinitesimal generator of $\{X_t\}_{t\in[t_0,T]}$. 
When $X_t = rt + L_t$, if $\{L_t\}_{t\in[t_0,T]}$ is a Brownian motion with triplet $(b,\sigma,0)$, using the condition (\ref{martingale_b}) the infinitesimal generator is:
\begin{align}\label{infinitesimal_BS}
 \LL^{BS} C(t,x) = \; \bigl( r - \frac{1}{2}\sigma^2 \bigr) \frac{\partial C(t,x)}{\partial x}
+ \frac{1}{2} \sigma^2 \frac{\partial^2  C(t,x)}{\partial x^2},
\end{align}
and the Eq. (\ref{PIDE}) becomes the well known \emph{Black-Scholes PDE}. 
When $\{L_t\}_{t\in[t_0,T]}$ is a Merton or VG process, the associated infinitesimal generators are 
(\ref{infinitesimal_M}) and  (\ref{VG_inf_gen}) respectively, and 
we obtain the \emph{Merton} and \emph{Variance Gamma PIDEs}
\footnote{In the generator formulas, using (\ref{mu}) and (\ref{martingale_b}), the parameter $\mu$ is replaced by $r$.}. 
The value of the call option is the solution of the PIDE (\ref{PIDE}) with the usual boundary conditions:
\begin{itemize}
 \item Payoff: $ C(T,x) = \max\{ e^x - K , 0 \} .$
 \item Lateral conditions: 
 $ C(t,x) \underset{x \to -\infty}{=} 0 \hspace{1em} \mbox{and} \hspace{1em} C(t,x) \underset{x \to \infty}{\sim} e^x - K e^{-r(T-t)} . $
\end{itemize}

In Section \ref{numerical} we solve the PIDEs with the above boundary conditions using the Implicit-Explicit method proposed in \cite{CoVo05b}.

\section{Properties of the Markov chain} \label{B}

\subsection{Transition probabilities}\label{B1}
Let us indicate the transition probabilities of $\{X_n\}_{n\in \N}$ by: 
\begin{equation}
 p(x_{i},x_{j}) := \PP(X_{n+1} = x_{j} | X_n = x_{i}). 
\end{equation}
The number of jumps of a jump-diffusion process is Poisson distributed, i.e. $N_t \sim \mbox{Po}(\lambda t)$, with $\lambda >0$.
In a small interval $\Delta t$, we can compute the first order approximated probabilities:
\begin{itemize}
 \item $\PP( N_{t+\Delta t} - N_t = 0) \overset{d}{=} \PP( N_{\Delta t} = 0) \approx 1-\lambda \Delta t $,
 \item $\PP( N_{\Delta t} = 1) = \PP( N_{\Delta t} > 0) \approx \lambda \Delta t $.
\end{itemize}
Let us consider the discrete dynamics of $\{X_n\}_{n\in \N}$ in Eq. (\ref{log_sde_discr}). 
We assume that in a small time step $\Delta t$ the process jumps exactly once ($N_{\Delta t} = 1$), or does not jump at all ($N_{\Delta t} = 0$).
The two possible mutually exclusive events are:
\begin{itemize}
 \item \textbf{Diffusion}. 
 The transition probability is $p^D(x_i, x_i + \Delta \Xi)$ and $\Delta \Xi \in \{ -h_x,0,h_x \}$.
 $p^D(x_i, x_{i+k}) = 0$ for $k \not \in \{-1,0,+1\}$. 
 \item \textbf{Jumps}. 
 The transition probability is $p^J(x_i, x_i + \Delta \tilde J)$. The random variable $\Delta \tilde J$ takes values 
 in $\Sigma_x$ (or $\Sigma^{\epsilon}_x$).
\end{itemize}
By conditioning on the values of $N_{\Delta t}$, the total transition probability is  
\begin{align}
 p(x_{i},x_{j}) &= p^D(x_{i},x_{j}) \, \PP( N_{\Delta t} = 0) + p^J(x_{i},x_{j}) \, \PP( N_{\Delta t} = 1) \\ \nonumber
	&= (1-\lambda \Delta t) \, p^D(x_{i},x_{j}) + (\lambda \Delta t ) \, p^J(x_{i},x_{j}).
\end{align}
The request of a positive probability impose a restriction on the time step size $\Delta t \leq \frac{1}{\lambda}$.
In this section we showed that the transition probability of $\{X_n\}_{n\in \N}$ is a convex combination of $p^D$ and $p^J$, 
as required by the property (\ref{local1}).
We refer to Chapter 5.6 of \cite{Kushner} for more details.

\subsection{Infinitesimal generator discretization}\label{B2}

In this section we provide an explicit form for the 
transition probabilities. This can be achieved by discretizing the infinitesimal generator of the process $\{X_t\}_{t\in[t_0,T]}$ in (\ref{portfolio_dynamics2}),
which corresponds to the first term inside the ``min'' in the HJB Equation (\ref{HJB2}). 
In the following steps we consider only the finite activity case, but the same idea works for an infinite activity process after using the Brownian approximation.  
In fact, the only difference between (\ref{log_sde_Merton}) and (\ref{log_sde_inf_act}) is the truncation in the integral.

In this section we drop the variable $y_j$ from $Q(t_n,y_j,x_i)$, because we are interested only in the uncontrolled log-price dynamics.
Let us assume for convenience that $Q$ is smooth enough, the derivatives are discretized by the finite differences:
\begin{itemize}
 \item Backward approximation in time: 
 $ \frac{\partial Q}{\partial t} \approx \frac{Q^{n+1}_{i} - Q^{n}_{i}}{\Delta t} $.
 \item Central approximation in space: 
 $ \frac{\partial Q}{\partial x} \approx \frac{Q^{n+1}_{i+1} - Q^{n+1}_{i-1}}{2 h_x} $.
 \item Second order in space:   
 $ \frac{\partial^2 Q}{\partial x^2} \approx \frac{Q^{n+1}_{i+1} + Q^{n+1}_{i-1} -2 Q^{n+1}_{i}}{h_x^2} $.
\end{itemize}
The integral terms in (\ref{HJB2}) are truncated and restricted to the domain 
$ \bigl[-B_1,B_2\bigr] = \bigl[ ( -K_1-1/2 )h_x , ( K_2+1/2 )h_x \bigr] $\footnote{If the integral has a truncation parameter as in (\ref{VG_inf_gen}), 
we choose $\epsilon = 1.5h_x$ and the 
restricted domain becomes $ \bigl[-B_1,-\epsilon \bigr]\bigcup \bigl[\epsilon,B_2 \bigr] = \bigl[ ( -K_1-1/2 )h_x , -3/2 h_x \bigr] \bigcup \bigl[ 3/2 h_x, ( K_2+1/2 )h_x \bigr] $. }.
The discretization is obtained by approximating the integral by Riemann sums (see \cite{CoVo05b}):
\begin{equation}\label{trap_quad}
  \int_{-B_1}^{B_2}  Q(t_{n+1},y_j,x_i +z) \nu(dz) \approx \sum_{k = -K_1}^{K_2} \nu_k Q^{n+1}_{i+k}, 
\end{equation}
where
\begin{equation}\label{nu1}
 \nu_k = \int_{(k-\frac{1}{2}) h_x}^{(k+\frac{1}{2}) h_x} \nu(z) dz, \hspace{1em} \mbox{ for } \hspace{1em} -K_1 \leq k \leq K_2. 
\end{equation}
We define the discrete version of $\hat m := \int_{-B_1}^{B_2} (e^z-1) \nu(dz)$, $\hat \lambda := \int_{-B_1}^{B_2} \nu(dz)$ 
and $\hat \alpha := \frac{1}{\hat \lambda} \int_{-B_1}^{B_2} z \nu(dz)$: 
\begin{equation}
 \hat m \approx \sum_{k = -K_1}^{K_2} (e^{kh_x}-1) \nu_k, \quad \;
 \hat \lambda \approx \sum_{k = -K_1}^{K_2} \nu_k,  \quad \;
 \hat \alpha \approx \frac{h_x}{\hat \lambda} \sum_{k = -K_1}^{K_2} k \nu_k.
\end{equation}
The jump transition probabilities can be defined as:
\begin{equation}\label{pJ}
 p^J_k := \frac{\nu_k}{\hat \lambda}. 
\end{equation}
The discretized equation becomes:
\begin{align}
&\frac{Q^{n+1}_{i} -Q^{n}_{i}}{\Delta t} + 
(\mu-\frac{1}{2}\sigma^2 - \hat m) \frac{Q^{n+1}_{i+1} -Q^{n+1}_{i-1}}{ 2 h_x} \\ \nonumber
&+ \frac{1}{2} \sigma^2 \frac{Q^{n+1}_{i+1} + Q^{n+1}_{i-1} - 2 Q^{n+1}_{i}}{h_x^2} 
 + \sum_{k = -K_1}^{K_2} \nu_k Q^{n+1}_{i+k} - \hat \lambda Q^{n}_i = 0.
\end{align}
Rearranging the terms we get: 
\begin{align*}
\biggl(1+ \hat \lambda \Delta t \biggr) Q^{n}_{i} =& \; p^D_{-1} Q^{n+1}_{i-1} + p^D_{0} Q^{n+1}_{i} + p^D_{+1} Q^{n+1}_{i+1} \\
&+ (\hat \lambda \Delta t) \sum_{k = -K_1}^{K_2} p^J_k Q^{n+1}_{i+k}.
\end{align*}
where we defined:
\begin{align}\label{pD} \nonumber
 p^D_{-1} &:= \bigl( -(\mu-\frac{1}{2}\sigma^2 -\hat m)\frac{\Delta t}{2 h_x} + \frac{1}{2}\sigma^2 \frac{\Delta t}{h_x^2}  \bigr) \geq 0 \\ 
 p^D_{0} &:= \bigl( 1 - \sigma^2 \frac{\Delta t}{h_x^2} \bigr) \geq 0 \\ \nonumber 
 p^D_{+1} &:= \bigl( (\mu-\frac{1}{2}\sigma^2 -\hat m)\frac{\Delta t}{2 h_x} + \frac{1}{2}\sigma^2 \frac{\Delta t}{h_x^2}  \bigr) \geq 0 
\end{align}
and $p^D_k := 0$ for $k \not \in \{-1,0,+1\}$.
From $p^D_0$ we obtain an important restriction on the time step size: $\Delta t \leq \frac{h_x^2}{\sigma^2}$, while the condition obtained from $p^D_{-1}$ and $p^D_{+1}$ i.e. 
$h_x \leq \frac{\sigma^2}{|\mu-\frac{1}{2}\sigma^2 -\hat m|}$ is easily satisfied.
If we bring the term $\bigl(1+\hat \lambda \Delta t \bigr)$ on the right hand side and use the first order Taylor approximation 
$\bigl(1+\hat \lambda \Delta t \bigr)^{-1} \approx 1 - \hat \lambda \Delta t$, we obtain: 
\begin{align}\label{expectation_tot}
 Q^{n}_{i} &\approx \bigl(1 - \hat \lambda \Delta t \bigr) \sum_{k=-1}^1 p^D_k \, Q^{n+1}_{i+k} 
           + \bigl( \hat \lambda \Delta t \bigr) \sum_{k = -K_1}^{K_2} p^J_k \, Q^{n+1}_{i+k} \\ 
           &= \sum_{k = -K_1}^{K_2} p_k \; Q^{n+1}_{i+k} \nonumber
\end{align}
where 
\begin{equation}\label{pK}
p_k = (1 - \hat \lambda \Delta t) p^D_k + ( \hat \lambda \Delta t ) p^J_k 
\end{equation}
is the total transition
probability, written in the form (\ref{local1}). It is straightforward to check that $\sum_k p_k =1$.  
Let us check that also the \emph{local consistency} conditions (\ref{local2}) are satisfied at first order in $\Delta t$:
\begin{align*}
\E \bigl[ \Delta X_n \bigr] =& \bigl(1 - \hat \lambda \Delta t \bigr) \sum_{k=-1}^1 p^D_k \, k h_x 
           + \bigl( \hat \lambda \Delta t \bigr) \sum_{k = -K_1}^{K_2} p^J_k \, k h_x \\
           =& \bigl(1 - \hat \lambda \Delta t \bigr) \bigl( \mu - \frac{1}{2}\sigma^2 -\hat m \bigr) \, \Delta t  
           + \bigl(\hat \lambda \Delta t \bigr) \, \hat \alpha \\
           \approx& \bigl( \mu - \frac{1}{2}\sigma^2 - \hat m + \hat \lambda \hat \alpha \bigr) \, \Delta t ,
\end{align*}
\begin{align*}
 \E \biggl[ \bigl[ \Delta X_n \bigr]^2 \biggr] =&
 \bigl(1 - \hat \lambda \Delta t \bigr) \sum_{k=-1}^1 p^D_k \, (k h_x)^2 + \bigl( \hat \lambda \Delta t \bigr) \sum_{k = -K_1}^{K_2} p^J_k \, (k h_x)^2 \\
           =& \bigl(1 - \hat \lambda \Delta t \bigr) \sigma^2 \, \Delta t  
           + \bigl(\hat \lambda \Delta t \bigr) \,  \hat \eta^2 \\
 \approx& \bigl( \sigma^2 + \hat \lambda \hat \eta^2 \bigr) \, \Delta t.
\end{align*}
We introduced $\hat \eta^2 := \frac{1}{\hat \lambda} \int_{-B_1}^{B_2} z^2 \nu(dz)  
\approx \frac{h_x^2}{\hat \lambda} \sum_{k = -K_1}^{K_2} k^2 \nu_k $, and indicate $\eta^2 := \frac{1}{\lambda} \int_{\R} z^2 \nu(dz) 
= \frac{1}{\lambda} \E \bigl[ \bigl| \int_{\R} z \tilde N(dt,dz) \bigr|^2 \bigr]$.
The discrete moments match the continuous moments when 
$h_x \to 0$ and $K_1,K_2 \to \infty$ such that 
$\hat \lambda \to \lambda$,
$\hat \alpha \to \alpha$ and $\hat \eta \to \eta$.

\section{Proofs of Theorems \ref{theorem1} and \ref{theorem2}}\label{C}  

Let us indicate the Eq. (\ref{HJB2}) by:
\begin{equation}\label{HJB4}
 F\bigl(\xx, Q(\xx), DQ(\xx), D^2Q(\xx), I(\xx,Q) \bigr) = 0,
\end{equation}
where $I(\xx,Q)$ is the integral operator and $\xx := (t,y,x)$. We assume that $Q \in C^{0}\bigl([t_0,T] \times \R^2\bigr)$ (see Section \ref{Section2.5}), and following
\cite{Kab16} we introduce the definition of \textbf{viscosity solution}. 
\begin{definition}
A function $Q \in C^{0}\bigl([t_0,T] \times \R^2\bigr)$ is a viscosity subsolution (supersolution) of \ref{HJB4} if for any $\bar \xx \in [t_0,T] \times \R^2$  
and any test function $\phi \in C^{1,1,2}\bigl([t_0,T] \times \R^2\bigr) \bigcap C_2\bigl([t_0,T] \times \R^2\bigr)$ such that $\phi(\bar \xx) = Q(\bar \xx)$ 
and such that $\bar \xx$ is a global 
maximum (minimum) of $Q-\phi$, the following is satisfied:
\begin{equation}
 F\bigl(\bar \xx, \phi(\bar \xx), D\phi(\bar \xx), D^2\phi(\bar \xx), I( \bar \xx,\phi) \bigr) 
 \underset{(\geq)}{\leq} 0. 
\end{equation}
A function $Q \in C^{0}\bigl([t_0,T] \times \R^2\bigr)$ is a viscosity solution of \ref{HJB2} if it is both a subsolution and a supersolution.
\end{definition}
  
Let us recall the statements of Theorems [\ref{theorem1}] and [\ref{theorem2}] and present their proofs.
\begin{teorema}[\textbf{\ref{theorem1}}]
 The Scheme [\ref{scheme}], with $p_k$ given by \ref{pK}, is monotone, stable and consistent. 
\end{teorema}
Let us prove the three properties separately.\\
\noindent
With the square brackets, as in $[\varepsilon^n_{j,i}]$, we indicate all the possible values $\varepsilon^{n'}_{j',i'}$ such that $(n',j',i')\not=(n,j,i)$. \\
The scheme is \textbf{monotone} i.e. for all $[\varepsilon^n_{j,i}] \geq 0$, it holds
$$\mathbb{S}\bigl(\rho, (n,j,i), Q^{n}_{j,i}, [Q^{n}_{j,i}] + [\varepsilon^n_{j,i}] \bigr) \leq \mathbb{S}\bigl( \rho, (n,j,i), Q^{n}_{j,i}, [Q^{n}_{j,i}] \bigr).$$ 
\begin{proof}
Let us write the Scheme [\ref{scheme}] as
$$ \mathbb{S} = 
  Q^{n}_{j,i} - \min \biggl\{ \sum_{k = -K_1}^{K_2} p_k \; Q^{n+1}_{j,i+k} , \, \min_{l} F(x_i,l,n) Q^n_{j+l,i}, \, \min_{m} G(x_i,m,n) Q^n_{j-m,i}  \biggr\} $$
Since $F(x_i,l,n) > 0$, $G(x_i,m,n) > 0$ for all $x_i, l, m, n$, and $p_k\geq 0$ for all $k$, the scheme $\mathbb{S}$ is a decreasing function of $[Q^{n}_{j,i}]$.
\end{proof}
\noindent
The scheme is \textbf{stable} i.e. for any $\rho>0$ there exists a bounded solution $Q^{\rho}$, with bound independent of $\rho.$ This is equivalent to prove that 
$||Q^n||_{\infty} \leq C$, for any $0\leq n \leq N$ and for $C$ independent on $\rho$.
\begin{proof}
 The terminal conditions (\ref{terminal_c}), (\ref{terminal_w}), (\ref{terminal_b}) are positive bounded functions in a bounded domain.
 Therefore we can write $0 \leq Q^N_{j,i} \leq C$ for all $i,j$, and $C$ does not depend on $\rho$. 
 Since all the coefficients are non-negative, it follows that the scheme is \emph{sign preserving} i.e.
 $Q^n_{j,i} \geq 0$ for all $i,j,n$. We can write:	
\begin{align*}
   Q^{n}_{j,i} &= \min \biggl\{ \sum_{k = -K_1}^{K_2} p_k \; Q^{n+1}_{j,i+k} , \, \min_{l} F(x_i,l,n) Q^n_{j+l,i}, \, \min_{m} G(x_i,m,n) Q^n_{j-m,i}  \biggr\} \\
	       &\leq \sum_{k = -K_1}^{K_2} p_k \; Q^{n+1}_{j,i+k} \; \leq \; || Q^{n+1} ||_{\infty}.    
\end{align*}
This holds for all $i,j$, then $ || Q^{n} ||_{\infty} \leq || Q^{n+1} ||_{\infty} $. Iterating we obtain: 
$$ || Q^{n} ||_{\infty} \leq || Q^{N} ||_{\infty} \leq C. $$
\end{proof}
\noindent
The scheme is \textbf{consistent} i.e. for any smooth function $\phi$
$$\mathbb{S} \bigl( \rho, \xx_{\rho}, \phi^{\rho}(\xx_{\rho}), [\phi^{\rho}]_{\xx_{\rho}} \bigr) \underset{\xx_{\rho}\to \xx}{\underset{\rho \to 0}{\longrightarrow}} 
F\bigl(\xx, \phi(\xx), D\phi(\xx), D^2\phi(\xx), I(\xx,\phi) \bigr).$$
with $\xx_{\rho} := (t_n,y_j,x_i)$.
\begin{proof}
 Now we look at the following cases, corresponding to each minimum value in the Scheme [\ref{scheme}]: \\
\noindent 
1) For some $l>0$, it holds
  \begin{align*}
   0 &= e^{\bigl(\frac{\gamma}{\delta(t_n,T)} (1+\theta_b) e^{x_i} lh_y \bigr)}  \phi \bigl( t_n,y_j + lh_y,x_i\bigr) - \phi \bigl(t_n,y_j,x_i\bigr) \\
     &= \biggl( 1 + \frac{\gamma (1+\theta_b) e^{x_i}}{\delta(t_n,T)} lh_y + \mathcal{O}(h_y) \biggr) 
     \biggl( \phi(t_n,y_j,x_i) + \frac{\partial \phi}{\partial y}\bigg|_{y_j} l h_y + \mathcal{O}(h_y) \biggr) 
          - \phi(t_n,y_j,x_i) \\
     &= \frac{\partial \phi}{\partial y}\bigl(t_n,y_j,x_i\bigr)  + \frac{\gamma}{\delta(t_n,T)} (1+\theta_b) e^{x_i} \, \phi \bigl(t_n,y_j,x_i\bigr) + \mathcal{O}(h_y).     
  \end{align*}
  \noindent 
  2) For some $m>0$, an analogous computation leads to  
  $$ - \frac{\partial \phi}{\partial y}\bigl(t_n,y_j,x_i\bigr)  - \frac{\gamma}{\delta(t_n,T)} (1-\theta_s) e^{x_i} \, \phi \bigl(t_n,y_j,x_i\bigr) + \mathcal{O}(h_y) = 0. $$
  \noindent 
  3) When $\sum_{k = -K_1}^{K_2} p_k \; \phi(t_{n+1},y_j,x_{i+k}) - \phi(t_n,y_j,x_i) = 0$ let us consider the expression (\ref{pK}), and expand $p^D$ and
  $p^J$ using (\ref{pD}) and (\ref{pJ}):
  \begin{align*}
   & (1-\sigma^2\frac{\Delta t}{h_x^2}) \biggl( \phi + \frac{\partial \phi}{\partial t}\bigg|_{t_n} \Delta t + \mathcal{O}(\Delta t^2) \biggr)  
   -\hat \lambda \Delta t \phi(t_{n+1},y_j,x_i) - \phi(t_n,y_j,x_i) \\
   &+ \biggl( (\mu -\frac{1}{2}\sigma^2 -\hat m) \frac{\Delta t}{2h_x} + \frac{1}{2}\sigma^2\frac{\Delta t}{h_x^2} + \mathcal{O}(\Delta t^2) \biggr) 
   \bigl( \phi + \frac{\partial \phi}{\partial x}\bigg|_{x_i} h_x + \frac{1}{2} \frac{\partial^2 \phi}{\partial x^2} h_x^2 +\mathcal{O}(h_x^3) \biggr) \\
   &+ \biggl( -(\mu -\frac{1}{2}\sigma^2 -\hat m) \frac{\Delta t}{2h_x} + \frac{1}{2}\sigma^2\frac{\Delta t}{h_x^2} + \mathcal{O}(\Delta t^2) \biggr) 
   \bigl( \phi - \frac{\partial \phi}{\partial x}\bigg|_{x_i} h_x + \frac{1}{2} \frac{\partial^2 \phi}{\partial x^2} h_x^2 +\mathcal{O}(h_x^3) \biggr) \\
   &+ \hat \lambda \Delta t \sum_{k = -K_1}^{K_2} \frac{\nu_k}{\hat \lambda} \; \phi(t_{n+1},y_j,x_{i+k})  = 0. \\
   \end{align*}
Let us replace all the terms at $t_{n+1}$ by using the first order Taylor approximation 
$\phi(t_{n+1},\cdot,\cdot) = \phi(t_{n},\cdot,\cdot) + \frac{\partial \phi}{\partial t}\bigg|_{t_n} \Delta t + \mathcal{O}(\Delta t^2)$.
The two terms in $\hat \lambda$ can be rewritten as $ \Delta t \sum_{k = -K_1}^{K_2} \nu_k \; \bigl( \phi(t_{n},y_j,x_{i+k}) - \phi(t_{n},y_j,x_{i}) \bigr)$. Using the 
approximation (\ref{trap_quad}) and (\ref{nu1}) we obtain
\begin{align*}
  & \frac{\partial \phi}{\partial t} + (\mu -\frac{1}{2}\sigma^2 -\hat m) \frac{\partial \phi}{\partial x} + \frac{1}{2} \sigma^2 \frac{\partial^2 \phi}{\partial x^2} \\
  & + \int_{-B_1}^{B_2} \bigl( \phi(t_{n},y_j,x_{i}+z) - \phi(t_{n},y_j,x_{i}) \bigr) \nu(z) dz + \mathcal{O}(\Delta t) + \mathcal{O}(h_x) = 0.    
\end{align*}
When sending $\rho \to 0$, $\xx_{\rho} \to \xx$ and $B_1,B_2 \to \infty$ we obtain the desired result for all 1) 2) and 3).
\end{proof}

\begin{teorema}[\textbf{\ref{theorem2}}]
 The solution $Q^{\rho}$ of (\ref{scheme1}) converges uniformly to the unique viscosity solution of (\ref{HJB4}).
\end{teorema}
\noindent
The proof follows closely \cite{BaSo91}. 
\begin{proof}
We only prove the subsolution case, since the arguments for the supersolution are identical.
Let $\bar \xx$ the strict global maximum of $Q-\phi$ for some $\phi \in C^{1,1,2} \bigcap C_2$, and such that $Q(\bar \xx) = \phi(\bar \xx)$.
Then there exist sequences $\rho_n$ and $\xx_n$ such that for $n\to \infty$: \\
$\rho_n \to 0$, $\xx_n \to \bar \xx$, $Q^{\rho_n}(\xx_n) \to Q(\bar \xx)$ and $\xx_n$ is a global maximum of $Q^{\rho_n}(\cdot) - \phi(\cdot)$. 
Let us define $\xi_n := Q^{\rho_n}(\xx_n) - \phi(\xx_n)$, such that $\xi_n \to 0$ when $n\to \infty$. For any $\xx$ it holds $Q^{\rho_n}(\xx) \leq \phi(\xx) +\xi_n $.
Let us consider the Scheme [\ref{scheme}]:
\begin{align*}
 0 =& \; \mathbb{S} \bigl( \rho_n, \xx_n, Q^{\rho_ n}(\xx_n), [Q^{\rho_ n}]_{\xx_n} \bigr) \\
 \geq& \; \mathbb{S} \bigl( \rho_n, \xx_n, \phi(\xx_n) + \xi_n, [\phi + \xi_n]_{\xx_n} \bigr),
\end{align*}
where we used the monotonicity property. By sending $n\to\infty$ and thanks to the consistency property, we obtain: 
$$ F\bigl(\bar \xx, \phi(\bar \xx), D\phi(\bar \xx), D^2\phi(\bar \xx), I(\bar \xx,\phi) \bigr) \leq 0. $$
\end{proof}

\bibliographystyle{apalike}
\bibliography{bibliography_NC}

\begin{thebibliography}{}

\bibitem[Andersen and Andreasen, 2000]{AnAn00}
Andersen, L. and Andreasen, J. (2000).
\newblock Jump-diffusion processes: Volatility smile fitting and numerical
  methods for pricing.
\newblock {\em Rev. Derivatives Research}, 4:231 -- 262.

\bibitem[Applebaum, 2009]{Applebaum}
Applebaum, D. (2009).
\newblock {\em Lévy Processes and Stochastic Calculus}.
\newblock Cambridge University Press; 2nd edition.

\bibitem[Barles and Soner, 1998]{BaSo98}
Barles, G. and Soner, H.~M. (1998).
\newblock Option pricing with transaction costs and a nonlinear
  {B}lack-{S}choles equation.
\newblock {\em Finance and Stochastics}, 2(4):369--397.

\bibitem[Barles and Souganidis, 1991]{BaSo91}
Barles, G. and Souganidis, P. (1991).
\newblock Convergence of approximation schemes for fully nonlinear second order
  equations.
\newblock {\em J. Asymptotic Analysis}, 4:271--283.

\bibitem[Benth et~al., 2002]{Benth02}
Benth, F., Karlsen, K., and Reikvam, K. (2002).
\newblock Portfolio optimization in a {Lévy} market with intertemporal
  substitution and transaction costs.
\newblock {\em Stochastics and Stochastics Report}, 74(3-4):517--569.

\bibitem[Black and Scholes, 1973]{BS73}
Black, F. and Scholes, M. (1973).
\newblock The pricing of options and corporate liabilities.
\newblock {\em The Journal of Political Economy}, 81(3):637--654.

\bibitem[Carmona, 2009]{Carmona}
Carmona, R. (2009).
\newblock {\em Indifference Pricing}.
\newblock Princeton University Press.

\bibitem[Clewlow and Hodges, 1997]{ClHo97}
Clewlow, L. and Hodges, S. (1997).
\newblock Optimal delta hedging under transaction costs.
\newblock {\em Jornal of Economic Dynamics and Control}, 21:1353--1376.

\bibitem[Cont, 2001]{Cont01}
Cont, R. (2001).
\newblock Empirical properties of asset returns: stylized facts and statistical
  issues.
\newblock {\em Quantitative Finance}, 1(2):223--236.

\bibitem[Cont and Tankov, 2003]{Cont}
Cont, R. and Tankov, P. (2003).
\newblock {\em Financial Modelling with Jump Processes}.
\newblock Chapman and Hall/CRC; 1 edition.

\bibitem[Cont and Voltchkova, 2005]{CoVo05b}
Cont, R. and Voltchkova, E. (2005).
\newblock A finite difference scheme for option pricing in jump diffusion and
  exponential {L}\'evy models.
\newblock {\em SIAM Journal of numerical analysis}, 43(4):1596--1626.

\bibitem[Damgaard, 1998]{Damgaard}
Damgaard, A. (1998).
\newblock {\em Optimal Portfolio Choice and Utility Based Option Pricing in
  Markets with Transaction Costs}.
\newblock PhD thesis, School of Business and Economics, Odense University.

\bibitem[Davis and Panas, 1994]{DaPa94}
Davis, M. H.~A. and Panas, V.~G. (1994).
\newblock The writing price of a {E}uropean contingent claim under proportional
  transaction costs.
\newblock {\em Computational and Applied Mathematics}, 13(2):0101--8205.

\bibitem[Davis et~al., 1993]{DaPaZa93}
Davis, M. H.~A., Panas, V.~G., and Zariphopoulou, T. (1993).
\newblock {E}uropean option pricing with transaction costs.
\newblock {\em SIAM J. Control Optim.}, 31(2):470--493.

\bibitem[De~Vallière et~al., 2016]{Kab16}
De~Vallière, D., Kabanov, Y., and Lépinette, E. (2016).
\newblock Consumption-investment problem with transaction costs for
  lévy-driven price processes.
\newblock {\em Finance and Stochastics}, 20(3):705--740.

\bibitem[Fleming and Soner, 2005]{FlemingSoner}
Fleming, W.~H. and Soner, M.~H. (2005).
\newblock {\em Controlled Markov Processes and Viscosity Solutions}.
\newblock Springer; 2nd edition.

\bibitem[Florescu et~al., 2014]{FlMaSe14}
Florescu, I., Mariani, M.~C., and Sengupta, I. (2014).
\newblock Option pricing with transaction costs and stochastic volatility.
\newblock {\em Electronic Journal of Differential Equations}, 2014(165):1--19.

\bibitem[Forsyth and Huang, 2012a]{FoHu12b}
Forsyth, P. and Huang, Y. (2012a).
\newblock Analysis of a penalty method for pricing a guaranteed minimum
  withdrawal benefit (gmwb).
\newblock {\em IMA, Journal of Numerical Analysis}, 32:320--351.

\bibitem[Forsyth and Huang, 2012b]{FoHu12a}
Forsyth, P. and Huang, Y. (2012b).
\newblock Iterative methods for solution of the singular control formulation of
  a gmwb pricing problem.
\newblock {\em Numerische Mathematik}, 122:133--167.

\bibitem[Framstad et~al., 1999]{OkSu01}
Framstad, N., \O{}ksendal, B., and Sulem, A. (1999).
\newblock Optimal consumption and portfolio in a jump diffusion market with
  proportional transaction costs.
\newblock {\em Journal of Mathematical Economics}, 35:233--257.

\bibitem[Grinold and Kahn, 1999]{GK99}
Grinold, R.~C. and Kahn, R.~N. (1999).
\newblock {\em Active Portfolio Management}.
\newblock McGraw-Hill Education; 2 edition.

\bibitem[Haentjens, 2013]{Haentjens13}
Haentjens, T. (2013).
\newblock Efficient and stable numerical solution of the
  heston–cox–ingersoll–ross partial differential equation by alternating
  direction implicit finite difference schemes.
\newblock {\em International Journal of Computer Mathematics},
  90(11):2409--2430.

\bibitem[Hodges and Neuberger, 1989]{HoNe89}
Hodges, S.~D. and Neuberger, A. (1989).
\newblock Optimal replication of contingent claims under transaction costs.
\newblock {\em The Review of Future Markets}, 8(2):222–239.

\bibitem[Kushner and Dupuis, 2001]{Kushner}
Kushner, H. and Dupuis, P.~G. (2001).
\newblock {\em Numerical Methods for Stochastic Control Problems in Continuous
  Time}.
\newblock Springer; 2nd ed.

\bibitem[Kushner and Martins, 1991]{MaKu91}
Kushner, H.~J. and Martins, F.~L. (1991).
\newblock Numerical methods for stochastic singular control problems.
\newblock {\em SIAM Journal of Control and Optimization}, 29(6):1443--1475.

\bibitem[Leland, 1985]{Le85}
Leland, H. (1985).
\newblock Option pricing and replication with transaction costs.
\newblock {\em The Journal of Finance}, 40(5):1283--1301.

\bibitem[Madan et~al., 1998]{MCC98}
Madan, D., Carr, P., and Chang, E. (1998).
\newblock The {V}ariance {G}amma process and option pricing.
\newblock {\em European Finance Review}, 2:79–105.

\bibitem[Madan and Seneta, 1990]{MaSe90}
Madan, D. and Seneta, E. (1990).
\newblock The {V}ariance {G}amma {(V.G.)} model for share market returns.
\newblock {\em The journal of Business}, 63(4):511--524.

\bibitem[Merton, 1976]{Me76}
Merton, R. (1976).
\newblock Option pricing when underlying stock returns are discontinuous.
\newblock {\em Journal of Financial Economics}, 3:125--144.

\bibitem[Mocioalca, 2007]{Mocio07}
Mocioalca, O. (2007).
\newblock Jump diffusion options with transaction costs.
\newblock {\em Revue Roumaine de Mathematiques Pures et Appliquees},
  52(3):349--366.

\bibitem[Monoyios, 2003]{Mon03}
Monoyios, M. (2003).
\newblock Efficient option pricing with transaction costs.
\newblock {\em Journal of Computational Finance}, 7:107--128.

\bibitem[Monoyios, 2004]{Mon04}
Monoyios, M. (2004).
\newblock Option pricing with transaction costs using a {M}arkov chain
  approximation.
\newblock {\em Jornal of Economic Dynamics and Control}, 28:889--913.

\bibitem[Sato, 1999]{Sato}
Sato, K.~I. (1999).
\newblock {\em Lévy processes and infinitely divisible distributions}.
\newblock Cambridge University Press.

\bibitem[Sengupta, 2014]{Sengu14}
Sengupta, I. (2014).
\newblock Option pricing with transaction costs and stochastic interest rate.
\newblock {\em Applied Mathematical Finance}, 21(5):399--416.

\bibitem[Wang and Li, 2014]{Song14}
Wang, S. and Li, W. (2014).
\newblock A numerical method for pricing european options with proportional
  transaction costs.
\newblock {\em Journal of Global Optimization}, 60(1):59--78.

\bibitem[Whalley and Wilmott, 1997]{WhWi97}
Whalley, A.~E. and Wilmott, P. (1997).
\newblock An asymptotic analysis of an optimal hedging model for option pricing
  with transaction costs.
\newblock {\em Mathematical Finance}, 7(3):307--324.

\end{thebibliography}

\end{document}